\documentclass[11pt]{article}
\usepackage{jcappub,natbib}
\usepackage{amsmath}    % required for `\align' (yatex added)
\newcommand{\nnmb}{\nonumber\\}
\newcommand{\braket}[3]{\left<{#1}\left|{#2}\right|{#3}\right>}
\newcommand{\im}{\mathrm{Im}\,}
\newcommand{\re}{\mathrm{Re}\,}
\newcommand{\ds}{\displaystyle}
{\baselineskip0pt
\rightline{\baselineskip16pt\rm\vbox to-20pt{
            \hbox{YITP-13-85}
\vss}}%
}
\title{Bispectrum from open inflation}
\author[a]{Kazuyuki Sugimura,}
\author[b,c,d]{Eiichiro Komatsu}
\affiliation[a]{Yukawa Institute for Theoretical Physics, Kyoto
University, Kyoto, Japan}
\affiliation[b]{Max-Planck-Institut f\"ur Astrophysik, Karl-Schwarzschild Str. 1, 85741 Garching, Germany}
\affiliation[c]{Kavli Institute for the Physics and Mathematics of the
Universe, Todai Institutes for Advanced Study, the University of Tokyo,
Kashiwa, Japan 277-8583 (Kavli IPMU, WPI)} 
\affiliation[d]{Texas Cosmology Center and the Department of Astronomy,
The University of Texas at Austin, 1 University Station, C1400, Austin,
TX 78712, USA} 
\emailAdd{sugimura@yukawa.kyoto-u.ac.jp}
\emailAdd{komatsu@mpa-garching.mpg.de}
\abstract{%
We calculate the bispectrum of primordial curvature perturbations,
$\zeta$, generated during ``open inflation.'' Inflation occurs inside a
bubble nucleated via quantum tunneling from the background false vacuum
state. Our universe lives inside the bubble, which can be described as a
 Friedman-Lema\^itre-Robertson-Walker 
 (FLRW) universe with negative spatial curvature, undergoing
 slow-roll inflation. We pay special attention to the issue of
 an initial state for quantum fluctuations.
 A ``vacuum state'' defined by a positive-frequency mode in de
 Sitter space charted by open coordinates is different from the
 Euclidean vacuum (which is equivalent to the so-called ``Bunch-Davies vacuum''
 defined by a positive-frequency mode in de Sitter space charted by flat
 coordinates). Quantum tunneling (bubble nucleation) then modifies the
 initial state away from the original Euclidean vacuum. While most of the
 previous study on modifications of the initial 
 quantum state introduces, by hand, an initial time at which the quantum
 state is modified as well as the form of the modification, an effective
 initial time naturally emerges and the form is fixed by
 quantum tunneling in open inflation models. Therefore, open inflation
 enables a 
 self-consistent computation of the effect of a modified initial state
 on the bispectrum.
 We find a term which goes as
 $\langle\zeta_{{\bf k}_1}\zeta_{{\bf k}_2}\zeta_{{\bf
 k}_3}\rangle\propto 1/{k_1^2k_3^4}$ 
 in the so-called squeezed configurations, $k_3\ll k_1\approx k_2$, in
 agreement with the previous study on modifications of the initial
 state. The bispectrum in the exact folded limit, e.g.,
 $k_1=k_2+k_3$, is also enhanced and remains finite.
However, these terms are exponentially suppressed when the
 wavelength of $\zeta$ is smaller than the curvature radius of the
 universe. 
 The leading-order bispectrum is equal to the
 usual one from single-field slow-roll inflation;
the terms specific for open inflation arise only in the sub-leading
order when the wavelength of $\zeta$ is smaller than the curvature radius.}
\begin{document}
\maketitle

%%%%%%%%%%%%%%%%%%%%%%%%%%%%%%%%%%%%%%
%%%%%%%%%%%%%%%%%%%%%%%%%%%%%%%%%%%%%%
\section{Introduction}
\label{sec:theme}
%%%%%%%%%%%%%%%%%%%%%%%%%%%%%%%%%%%%%%
%%%%%%%%%%%%%%%%%%%%%%%%%%%%%%%%%%%%%%

``Open inflation'' models \cite{Gott:1982zf,Gott:1984ps} offer
an attractive framework for understanding the origin of our
universe. According to this framework, our universe is contained within
a single bubble, which was nucleated from a surrounding false vacuum
state. This is attractive because there is no physical singularity at
the beginning of {\it our} universe, which is the moment of the bubble
nucleation in the de Sitter background. In the simplest scenario worked
out by Coleman and De Luccia 
\cite{Coleman:1980aw}, the metric inside the bubble is a homogeneous
and isotropic Friedman-Lema\^itre-Robertson-Walker (FLRW) metric with
negative spatial curvature. Therefore, the homogeneity and isotropy problems do
not exist in this scenario.\footnote{However, one may argue that this is
merely a consequence of the assumptions made in the analysis of Coleman and De
Luccia, who studied a bubble nucleation in a homogeneous and isotropic
background.} We still need inflation
\cite{Starobinsky:1980te,Sato:1980yn,Guth:1980zm,Linde:1981mu,Albrecht:1982wi}
to make geometry of the observable universe sufficiently flat, at the
level compatible with observations
\cite{Bennett:2012zja,Hinshaw:2012aka,Ade:2013zuv,Ade:2013uln}; hence the
term, ``open inflation.''

How can we test open inflation models? While this is an attractive
framework, is it testable/falsifiable?  An obvious observable is the
spatial curvature of the observable universe. Detection of negative
curvature would greatly strengthen the case for open inflation
models, while detection of positive curvature would rule them out
\cite{Freivogel:2005vv}. In any case, in order for us to have any access
to distinct signatures of open inflation, the total number of $e$-folds
of inflationary expansion must be close to the minimum that is required
to make geometry of the observable universe sufficiently flat. If so, we
may find 
signatures of open inflation in scalar \cite{Bucher:1994gb}
and tensor 
perturbations \cite{Bucher:1997xs,Tanaka:1997kq} in  
temperature anisotropy of the cosmic microwave background (CMB) 
\cite{Yamamoto:1995sw,Yamamoto:1996qq,Hu:1997ws,GarciaBellido:1997hy,Sasaki:1997ex,Linde:1999wv}.\footnote{See
refs.~\cite{Lyth:1990dh,Ratra:1994vw,Ratra:1994dm,Kamionkowski:1994sv,Gorski:1994zs}
for earlier calculations based upon a conformal vacuum state (rather than the
Euclidean vacuum state we shall work with in this paper).}
Another
possible observable is a signature of other bubbles colliding with ours 
\cite{Chang:2008gj,Czech:2010rg,Feeney:2010jj}.

\begin{figure}
\begin{center}
 \includegraphics[width=10cm]{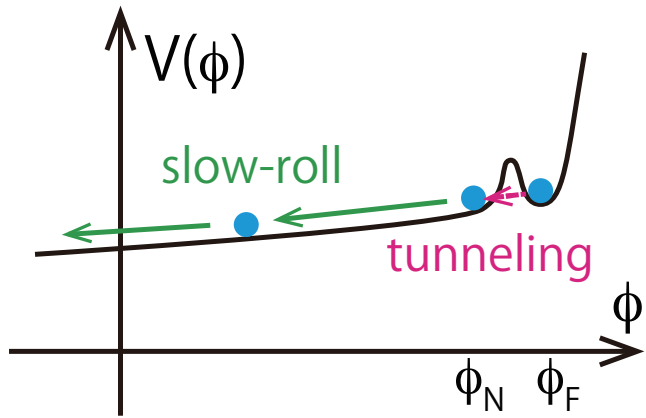} 
\end{center}
\caption{Potential for single-field open inflation.
Slow-roll inflation begins after a scalar field tunneling from the
 false vacuum state, $\phi_F$, to  the nucleation point,
 $\phi_N$. The essential features of this potential are: the potential
 heights before and after tunneling are  approximately the same,
 $V(\phi_F)\approx V(\phi_N)$; the barrier height is small compared to
 $V(\phi_F)$; and the barrier is narrow, $V\ll |d^2V/d\phi^2|$ at the barrier.}
\label{fig:pot}
\end{figure}

In this paper, we shall present the first computation of the
bispectrum from single-field open inflation.\footnote{See
ref.~\cite{Clunan:2009ib} for 
related work on the bispectrum from inflation with positive spatial
curvature.} Our model is based on ref.~\cite{Linde:1998iw}, and the
potential for the scalar (inflaton) field is shown in
figure~\ref{fig:pot}. In these models, a single scalar field is
responsible for both quantum tunneling and slow-roll inflation inside the
bubble. For simplicity, we 
shall ignore multi-field effects or the possibility of a rapid-roll era
soon after quantum tunneling, although these effects are potentially
significant \cite{Sugimura:2011tk,Yamauchi:2011qq}.
The bispectrum is the three-point function in Fourier space,
and we define it as $\langle\zeta_{{\bf k}_1}\zeta_{{\bf
k}_2}\zeta_{{\bf k}_3}\rangle=(2\pi)^3\delta_D(\sum_{i=1}^3{\bf
k}_i)B(k_1,k_2,k_3)$, where $\zeta_{\bf k}$ is the Fourier transform of
a curvature perturbation computed on the uniform density hypersurface.

Our work is motivated in part by recent work on the effects of a
modified initial state (non-Bunch-Davies initial state\footnote{We shall
define what we mean by the ``Bunch-Davies initial state'' in
section~\ref{sec:mode_open}.}) on the bispectrum
\cite{Chen:2006nt,Holman:2007na,Meerburg:2009ys,Agullo:2010ws,Ganc:2011dy,Chialva:2011hc,Kundu:2011sg,Dey:2011mj,Aravind:2013lra}. In
particular, it has been shown that a modified initial state yields an
enhanced bispectrum in the so-called squeezed limit, in which 
one of the wavenumbers is much less than the other two, i.e., $k_3\ll
k_1\approx k_2$. Specifically, $B(k,k,k_3)\propto
k_3^{-4}k^{-2}$ in the squeezed limit
\cite{Agullo:2010ws,Ganc:2011dy,Chialva:2011hc} instead of the usual
$B(k,k,k_3)\propto k_3^{-3}k^{-3}$ for single-field inflation models with a
Bunch-Davies initial state \cite{Maldacena:2002vr}. Such a distinct
dependence on $k_3$ has important implications for observations
\cite{Ganc:2012ae,Agullo:2012cs}. However, most of the previous
work on a modified initial state puts an initial state at some arbitrary
initial time by hand, without specifying its physical origin. The only
self-consistent computations that we are aware of use periodic features
in the scalar-field potential or kinetic term
\cite{Creminelli:2011rh,Chen:2010bka}.

In open inflation models, such a modified initial state naturally
emerges as a result of the evolution from the Euclidean vacuum (which is
equivalent to the Bunch-Davies vacuum in de Sitter space charted by flat
coordinates) via
quantum tunneling \cite{Yamamoto:1996qq,Garriga:1997wz,Garriga:1998he}.  
Therefore, open inflation enables a self-consistent computation of
the effect of a modified initial state on the bispectrum.

This paper is organized as follows. 
In section~\ref{sec:review}, we review open inflation and our coordinate system.
In section~\ref{sec:mode_open}, we review quantum field theory for a free
scalar field in open inflation.
In section~\ref{sec:qft}, we outline the in-in formalism on a Coleman-De
Luccia instanton background, with which we calculate the bispectrum in
section~\ref{sec:ng-gen}. 
Finally, we conclude in section~\ref{sec:conclusion}.
In appendix~\ref{sec:expansion}, we derive the second- and third-order
actions for scalar perturbations.
In appendix~\ref{sec:tunneling-region}, we show how we choose the paths
of integration when computing the correlation function.
In appendix~\ref{sec:tunneling-eval-each}, we show that all but one term
in the third-order action after field redefinition is important in the
sub-curvature approximation.
In appendix~\ref{sec:sque-limit-mode}, we show the correspondence
between the open  harmonics  and the Fourier modes in flat space in the
sub-curvature approximation.

We adopt the units $c=\hbar=8\pi G =1$.

%%%%%%%%%%%%%%%%%%%%%%%%%%%%%%%%%%%%%%
%%%%%%%%%%%%%%%%%%%%%%%%%%%%%%%%%%%%%%
\section{Open inflation}
\label{sec:review}
%%%%%%%%%%%%%%%%%%%%%%%%%%%%%%%%%%%%%%
%%%%%%%%%%%%%%%%%%%%%%%%%%%%%%%%%%%%%%
We describe quantum tunneling using
a Coleman-De Luccia (CDL) instanton, $\bar{\phi}(x)$ and $\bar{g}_{\mu
\nu}(x)$ \cite{Coleman:1980aw}. 
A CDL instanton, which is an Euclidean O(4)-symmetric solution, 
can be written as a function of only one variable, $\bar{\phi}(\tau)$
and $\bar{a}(\tau)$, where $\tau$ is the imaginary (Euclidean)
time. We study quantum tunneling using the Euclidean metric given by
\begin{equation}
ds^2=d\tau^2+\bar{a}^2(\tau)(d\chi^2+\sin^2\chi
d\Omega^2), 
\label{eq:euclidds}
\end{equation}
with $d\Omega^2$ being the metric of a 2-sphere ($S_2$). We then
analytically continue the solution beyond tunneling (i.e., bubble
nucleation) in order to describe the post-tunneling world.

The equations of motion (EOMs) for the instanton are given by
\begin{align}
 \frac{d^2\bar{\phi}(\tau)}{d\tau^2}+
3\frac{d\ln \bar{a}(\tau)}{d\tau}
\frac{d\bar{\phi}(\tau)}{d\tau}
-\frac{dV[\bar{\phi}(\tau)]}{d\bar{\phi}}=0&\,,\label{eq:66}\\
\left[\frac{d\ln \bar{a}(\tau)}{d\tau}\right]^2
= -\frac{V[\bar{\phi}(\tau)]}{3}+\frac{1}{\bar{a}^2(\tau)}&\,,
\label{eq:65}
\end{align}
where the potential, $V(\phi)$, has a false vacuum at $\phi_F$. We have
a slow-roll inflation phase following the quantum tunneling 
 to a nucleation point, $\phi_N$ (see
figure~\ref{fig:pot}).  

For simplicity, we shall assume that the potential barrier is small
compared with the vacuum energy of the false vacuum, and the
potential heights before and after tunneling as well as during the
subsequent slow-roll phase are approximately equal, i.e.,
$V(\phi_F)\approx V(\phi_N)\approx V(\phi_I)\equiv V_I$, where $\phi_I$
denotes the scalar field values during slow-roll inflation.
Then, we can approximate $V(\phi)$ as $V(\phi)\approx V_I$ in
eq.~\eqref{eq:65},
and $\bar{a}$ is approximately given by the scale factor of Euclidean de
Sitter spacetime,
\begin{align}
 \bar{a}(\tau)&=\frac{1}{H_I}\sin(H_I\tau),
\label{eq:46}
\end{align}
with $H_I^2=V_I/3$. 
With this choice of coordinates $\tau$, the two ends of the Euclidean
4-sphere in $\tau$ correspond to $H_I\tau=\pm \pi/2$. 
The boundary conditions for the instanton are
$d\bar{\phi}(\tau)/d\tau=0$ at $\tau=\pm \pi/(2H_I)$. 
We shall also assume that the potential barrier is
narrow, $H_I^2\ll |d^2V/d\phi^2|$. 

With these assumptions, $\bar{\phi}$ is approximately given
 by a thin-wall instanton solution:
\begin{align}
 \bar{\phi}(\tau)&\approx
\begin{cases}
\ds \phi_F&\ds \left(-\frac{\pi}{2H_I} \leq \tau < \frac{\pi}{2H_I}-R_W\right)
\\[10pt]
\ds \phi_N&\ds \left(\frac{\pi}{2H_I}-R_W< \tau \leq \frac{\pi}{2H_I}\right)
\end{cases}
\,,
\label{eq:5}
\end{align}
where $R_W$ is the radius of the bubble wall.

\begin{figure}
\begin{center}
 \includegraphics[width=10cm]{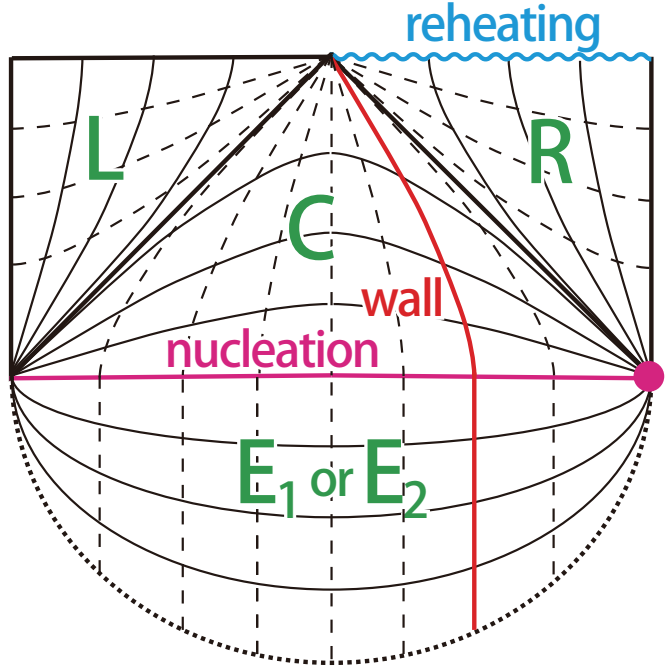}
\end{center}
\caption{Penrose-like diagram of open inflation. The top half
 (above the line labeled ``nucleation'') corresponds to the world
 after bubble nucleation, and the time is real. The bottom half
 corresponds to either one of the two hemispheres of the Euclidean
 4-sphere, $E_1$ (``north'') or $E_2$ (``south''), and
 the time is imaginary. (Also see figure~\ref{fig:inin_path}.)
 The solid and dashed lines show $r=\mbox{constant}$ and
 $t=\mbox{constant}$ lines, respectively, in the $R$-, $L$-, and
 $C$-regions, while they show $\chi=\mbox{constant}$ and
 $\tau=\mbox{constant}$, respectively, in the $E_1$- and $E_2$-regions.
 The $R$-region describes the 
 interior of the bubble, i.e., our universe. The $C$-region contains a
 worldline of the bubble wall, which separates the interior (located to
 the right of the line labeled ``wall'') and exterior of the bubble. The
 exterior of the bubble is in the false vacuum state with
 $\bar{\phi}=\phi_F$. The big filled circle at the right edge of the
 line labeled ``nucleation'' shows the location of the bubble center at
 the time of nucleation. Note that the bubble has a finite size upon
 nucleation. The $L$-region is in the false vacuum state with
 $\bar{\phi}=\phi_F$. The coordinate values are chosen  such that the
 right vertical edge of the $R$-region has $r_R=0$, and the thick solid
 line between the $R$- and $C$-regions have $t_R=0$ and $r_R=\infty$ as
 well as $t_C=\pi/(2H_I)$ and $r_C=\infty$. The line labeled
 ``nucleation'' has $r_C=0$ as well as $\chi=\pi/2$. The left vertical edge of
 the $L$-region 
 has $r_L=0$, and the thick solid line between the $L$- and $C$-regions
 have $t_L=0$ and $r_L=\infty$ as well as $t_C=-\pi/(2H_I)$ and
 $r_C=\infty$. Finally, while the top edge of the $L$-region has
 $t_L=\infty$, that of the $R$-region is somewhat fuzzy because
 slow-roll inflation in the $R$-region must end in a finite time and our
 universe must enter a radiation-dominated era after reheating (which is
 not shown in this diagram). While $t$ and $r$ in the $R$- and
 $L$-regions coincide with the familiar time and radial coordinates in a
 FLRW universe, those in the $C$-region do not. For example, a
 $t_C=\mbox{constant}$ line gives a worldline of a body with constant
 acceleration (which is analogous to the Rindler coordinates). Therefore,
 the bubble wall in the $C$-region is accelerated and its worldline
 approaches a null line as $r_C\to \infty$. In this paper, the
 wall worldline is along $t_C=R_W$  where $R_W$ is the radius of the
 bubble, and $\phi(t_C)=\phi_F$ and 
 $\phi_N$ for $-\pi/(2H_I)\le t_C<\pi/(2H_I)-R_W$ (exterior of the
 bubble) and $\pi/(2H_I)-R_W<t_C\le \pi/(2H_I)$ (interior), respectively
 (eq.~\eqref{eq:5}).
}
\label{fig:pen}
\end{figure}

The world after bubble nucleation can be described by analytical
continuation of a CDL instanton to Lorentzian regions.
For the Euclidean de Sitter spacetime defined by eq.~\eqref{eq:46},
analytical continuation is made with the following coordinate
system \cite{Sasaki:1994yt}, 
\begin{align}
\mbox{$E_1$- and $E_2$-regions:} 
& \begin{cases}
  \tau&\left(-\pi/(2H_I)\leq \tau\leq\pi/(2H_I)\right),\\
 \chi&\left(0\leq \chi\leq\pi\right),
 \end{cases}\label{eq:127-E}\\
\mbox{$R$-region:}
& \begin{cases}
  t_R= i(\tau-\pi/(2H_I))&\left(0\leq t_R<\infty\right),\\
 r_R=i\chi&\left(0\leq r_R<\infty\right),
 \end{cases}\label{eq:127-R}\\
\mbox{$L$-region:} 
& \begin{cases}
 t_L=i\left(-\tau-\pi/(2H_I)\right)&\left(0\leq t_L<\infty\right),\\
 r_L=i\chi&\left(0\leq r_L<\infty\right),
\end{cases}\label{eq:127-L}\\
\mbox{$C$-region:} 
& \begin{cases}
 t_C=\tau&\left(-\pi/(2H_I)\leq t_C\leq \pi/(2H_I)\right),\\
 r_C=i\left(\chi-\frac{\pi}{2}\right)&\left(0\leq r_C<\infty\right)\,.
\end{cases}
\label{eq:127-C}
\end{align}
The $E_1$- and $E_2$-regions are the original Euclidean regions
we have used to describe quantum tunneling,
which consist of 
two 4-hemispheres $E_1$ and $E_2$ defined by $0\leq\chi<\pi/2$
(``north'') and $\pi/2<\chi\leq\pi$ (``south''),
respectively. The others are the Lorentzian 
regions, which describe the world after quantum tunneling. Each
region and its physical meaning are illustrated in figure~\ref{fig:pen}.
The coordinates of a 2-sphere, $\Omega^i$, are commonly used for all regions.
The boundaries between the $C$- and $R$-regions and the $C$- and $L$-regions
are coordinate singularities.

With the coordinate system given in eq.~\eqref{eq:127-E} to \eqref{eq:127-C},
the metrics in the $E$-, $R$-, $L$-, and $C$-regions are given, respectively, by
\begin{align}
 ds^2
&=d\tau^2+H_I^{-2}\cos^2H_I\tau\left(d\chi^2+\sin^2\chi d\Omega^2\right)\nnmb
  &=-dt_R^2+H_I^{-2}\sinh^2H_It_R\left(dr_R^2+\sinh^2r_R d\Omega^2\right)\nnmb
&=-dt_L^2+H_I^{-2}\sinh^2H_It_L\left(dr_L^2+\sinh^2r_L d\Omega^2\right)\nnmb
&=dt_C^2+H_I^{-2}\cos^2H_It_C\left(-dr_C^2+\cosh^2r_C d\Omega^2\right)\,,
\label{eq:103}
\end{align}
and $\phi$'s in the corresponding regions are given, respectively, by
\begin{align}
\phi(\tau)&=\bar{\phi}(\tau)\,,\\
\phi(t_R)&=\bar{\phi}[-it_R+\pi/(2H_I)]\,, \\
\phi(t_L)&=\bar{\phi}[it_L-\pi/(2H_I)]\,,\\
\phi(t_C)&=\bar{\phi}(t_C)\,,
\end{align}
where $\bar{\phi}(\tau)$ is a solution to eqs.~\eqref{eq:66} and
\eqref{eq:65}.

We live in the future of the $R$-region, in which slow-roll inflation
occurs after quantum tunneling, and reheating follows after inflation (see
figure~\ref{fig:pen}). 
In the $R$-region, a $t_R={\rm const}$ surface is both a 3-hyperboloid 
and a $\phi={\rm const}$ surface; thus, we observe the $R$-region as a
FLRW universe with negative spatial curvature.

If we could solve eqs.~\eqref{eq:66} and \eqref{eq:65} exactly for
a given potential written by an analytical function, analytical
continuation given by eq.~\eqref{eq:127-R} 
to \eqref{eq:127-C} would give a complete description of the world after
quantum tunneling. However, since the solutions given in eqs.~\eqref{eq:46} and
\eqref{eq:5} are approximation, we
need to re-solve the EOM for $\phi$ and the Friedmann equation in a FLRW
universe with negative spatial curvature, in order to describe the
evolution of the universe in the region where the solutions given in
eqs.~\eqref{eq:46} and \eqref{eq:5} are not valid. We shall therefore
use eqs.~\eqref{eq:46} and \eqref{eq:5} as the initial conditions for
the subsequent evolution of $\phi(t)$ and $a(t)$.

For brevity, in the following we omit the subscript $R$ from $t_R$ and
$r_R$ unless otherwise stated. The metric in the $R$-region is given by
\begin{equation}
 ds^2=
-dt^2+a^2(t)\gamma_{ij} dx^idx^j\,,
\label{eq:7}
\end{equation}
where $\gamma_{ij}$ is the metric for the unit 3-hyperboloid given by
\begin{equation}
\gamma_{ij} dx^idx^j\equiv dr^2+\sinh^2r d\Omega^2\,,
\end{equation}
and $a(t)$ is the FLRW scale factor. 
The EOM for $\phi$ and the Friedmann equation are given, respectively,  by
\begin{align}
 \ddot{\phi}(t)+3H(t)\dot{\phi}(t)
+\frac{dV[\phi(t)]}{d\phi}= 0&\,,\label{eq:67}\\
H^2(t)
= \frac{V[\phi(t)]}{3}+\frac{1}{a^2(t)}&\,,
\label{eq:42}
\end{align}
where the over-dots denote derivatives with respect to $t$
and $H(t)\equiv \dot{a}(t)/a(t)$ is in general different from $H_I$.
The initial conditions are given 
by $a(0)=0$, $\dot{a}(0)=1$, $\phi(0)=\phi_N$, and $\dot{\phi}(0)=0$,
according to the CDL instanton solution.

At the onset of slow-roll inflation in the $R$-region, the universe is
still in a curvature-dominated era, i.e.,  
the spatial curvature term is still dominant in the Friedmann equation.
However, spatial curvature decays away as the universe expands,
and an inflationary era begins when the potential of $\phi$ becomes
dominant in the Friedmann equation. 
After reaching the inflationary era,
the universe evolves in the same way as the usual slow-roll inflation
scenario. We thus impose slow-roll conditions:
\begin{align}
 \epsilon&\equiv \frac{(dV/d\phi)^2}{2V^2}\ll 1\,,\qquad
 \epsilon_\eta\equiv \frac{(d^2V/d^2\phi)}{V}\ll 1\,.
\label{eq:43}
\end{align}
Although the slow-roll conditions may be broken
during the curvature-dominated era, we shall assume that the slow-roll
conditions are satisfied in the whole $R$-region.

Let us define the conformal time, $\eta\equiv \int_{\infty}^t dt'/a(t')$.
To the leading order of the slow-roll parameters, we have
$\eta\approx
-(1/2)\log\left((\cosh H_It+1)/(\cosh H_It-1)\right)$. 
The curvature-dominated era and the inflationary era
correspond approximately to $-\infty<\eta\lesssim-1$ and
$-1\lesssim\eta<0$, respectively. The scale factor and the Hubble
parameter are approximately given by 
\begin{align}
 a(\eta)&\approx 
\begin{cases}
\ds \frac{e^\eta}{2H_I}& \left(-\infty<\eta\lesssim -1\right)\\[8pt]
\ds-\frac{1}{H_I\eta}&\left(-1\lesssim \eta<0\right)
\end{cases}\,,\qquad
H(\eta)\approx
\begin{cases}
\ds -H_I\eta& \left(-\infty<\eta\lesssim -1\right)\\[5pt]
\ds H_I&\left(-1\lesssim \eta<0\right)
\end{cases}\,.
\label{eq:18}
\end{align}
The first and second derivatives of the scalar field with respect
to time are approximately given by
\begin{align}
\dot{\phi}(\eta)&\approx
\begin{cases}
\ds
-\frac{3\sqrt{2\epsilon}}{2}H_Ie^{\eta}
&\left(-\infty<\eta\lesssim -1\right)\\[5pt]
-\sqrt{2\epsilon}H_I&\left(-1\lesssim \eta<0\right)
\end{cases}
\,,\qquad
\ddot{\phi}(\eta)\approx
\begin{cases}
\ds
-\frac{3\sqrt{2\epsilon}}{4}H_I^2
&\left(-\infty<\eta\lesssim -1\right)\\[5pt]
\sqrt{2\epsilon}(\epsilon+\epsilon_\eta)H_I^2&\left(-1\lesssim \eta<0\right)
\end{cases}\,.
\label{eq:36}
\end{align}
We assume that the potential is a monotonically increasing
function of $\phi$ (except for the barrier), and thus $\dot{\phi}$ and
$\ddot{\phi}$ have definite signs.

%%%%%%%%%%%%%%%%%%%%%%%%%%%%%%%%%%
\section{Quantum field theory for a free scalar field in open inflation}
\label{sec:mode_open}
%%%%%%%%%%%%%%%%%%%%%%%%%%%%%%%%%%
In this section, we review quantum field theory (QFT) for a free scalar
field in open inflation \cite{Yamamoto:1996qq,Garriga:1997wz,Garriga:1998he}.
Hereafter, we shall consider only scalar-type perturbations.
We write the perturbed scalar field around the uniform background,
$\phi(t)$, as 
\begin{align}
 \phi(t,{\bf x})=\phi(t)+\varphi(t,{\bf x})\,,
\end{align}
where $\varphi$ is the scalar field perturbation. We write the spatial
coordinates as ${\bf x}\equiv(r,\Omega)$ for brevity. From now on, we
shall use $\phi$ to denote the background scalar field, $\phi(t)$,
unless otherwise stated. 

The perturbed metric in the $R$-region around the metric given by
eq.~\eqref{eq:7} is written in the Arnowitt-Deser-Misner (ADM)
form as \cite{Arnowitt:1962hi} 
\begin{align}
 ds^2=-N^2dt^2+h_{ij}\left(dx^i+N^idt\right)\left(dx^j+N^jdt\right)\,,
\label{eq:76}
\end{align}
where $h_{ij}=a^2(t)e^{2\zeta}\,\gamma_{ij}$ is the spatial metric
with the curvature perturbation $\zeta$,
$N$ the lapse function, and $N_i$ the shift vector.
We fix the gauge degrees of freedom by taking the uniform curvature (flat)
gauge, $\zeta\equiv 0$, and obtain the quadratic action for scalar
perturbations by substituting the constraint equations
into the original action. As we show in appendix~\ref{sec:second-order-action}, 
the quadratic action for scalar perturbations up to the leading order in
the slow-roll parameters is given by 
\begin{align}
  S_2&\left(\equiv\int d^4x \mathcal{L}_0\right)
=\int dt\,d^3{\bf x}\,a^3\sqrt{\gamma}
\left[
\frac{1}{2}\dot{\varphi^2}-\frac{1}{2a^2}\partial_i\varphi\partial^i\varphi
\right]\,,
\label{eq:47}
\end{align}
where $\mathcal{L}_0$ is the Lagrangian density for free theory,
$\partial_i$ denotes 3-dimensional covariant derivatives with respect to
$\gamma_{ij}$,  
and $\partial^i$ is defined by $\partial^i\equiv \gamma^{ij}\partial_j$.
eq.~\eqref{eq:47}, derived in the $R$-region,
can also be used outside the $R$-region by analytical continuation,
except near the bubble wall, for which a non-zero mass term should be
inserted into the action.

Next, we obtain the 2-point function of $\varphi$.
We promote $\varphi$ to an operator, $\hat{\varphi}$, and write
\begin{align}
 \hat{\varphi}(x)&=\sum_{\bf k}\left[
u_{\bf k}(x)\hat{a}_{\bf k}+v_{\bf k}(x)\hat{a}_{\bf k}^\dagger\right]\,,
\label{eq:57}
\end{align}
where $\hat{a}_{\bf k}$ and $\hat{a}_{\bf k}^\dagger$ 
are the annihilation and creation operators, respectively,
that satisfy the commutation relation,
$[\hat{a}_{\bf k}, \hat{a}_{{\bf k}'}^\dagger]=\delta_{{\bf k}{\bf
k}'}$; and $u_{\bf k}(x)$ and $v_{\bf k}(x)$ are mode functions, which
form a complete set.
The mode functions are solutions of the linearized EOMs.
Whilst $v_{\bf k}(x)=u^*_{\bf k}(x)$ in the Lorentzian region, the
same relation does not necessarily hold in the Euclidean region, as
the argument $x$ may contain an imaginary part. 

The choice of $u_{\bf k}$ selects a ``vacuum state,''
$\left|0\right>$, which is annihilated by the annihilation operator in
eq.~\eqref{eq:57}, $\hat{a}_{\bf k}\left|0\right>=0$. 
The most natural choice of a vacuum state in de Sitter space is
the Euclidean vacuum, $\left|0_E\right>$. The Euclidean vacuum is
defined by a purely 
positive-frequency function which is analytical in $E_2$. The two-point
function (Wightman function) computed with respect to this vacuum state,
$\left<0_E\right|\varphi(x)\varphi(x')\left|0_E\right>$, is
invariant under de Sitter group SO(4,1) (see section~5.4 of
ref.~\cite{birrell-davies}).\footnote{The Euclidean vacuum state is
equivalent to the Bunch-Davies vacuum state \cite{Bunch:1978yq}, which
is perhaps more familiar to cosmologists. The Bunch-Davies state is defined by a
positive-frequency function of de Sitter space charted by flat 
coordinates. Specifically, for flat coordinates of
$ds^2=a^2(\eta)(-d\eta^2+d{\bf x}^2)$ with $a(\eta)=-1/(H\eta)$, a
positive-frequency 
function for a minimally-coupled scalar field with mass $m$ is given by
the Hankel function of the first kind as $u_{\bf k}=\frac{\sqrt{-\pi\eta}}{2(2\pi)^{3/2}a}H_{\nu}^{(1)}(-k\eta)e^{i\bf{k}\cdot{\bf
x}}$ with $\nu^2=\frac94-\frac{m^2}{H^2}$ for $\eta<0$.
This function is analytical in the entire lower half complex $\eta$ plane
[$\im(\eta)<0$], and $u_{\bf k}$ is finite in the limit of
$\eta\to -(1+i\epsilon)\infty$. This procedure yields a purely
positive-frequency function of the Euclidean vacuum state.
The annihilation operator defined with respect to this $u_{\bf k}$
annihilates the Euclidean vacuum, $\hat{a}_{\bf k}\left|0_E\right>=0$,
and a comoving observer 
in de Sitter space charted by flat coordinates does not detect
particles  (see section~5.4 of
ref.~\cite{birrell-davies}; also see ref.~\cite{Allen:1985ux}).}

However, a positive-frequency function of de
Sitter space charted by open coordinates does not give a purely
positive-frequency function of the Euclidean vacuum state. This is not a
surprise: a positive-frequency function defined in a given coordination
is typically a mixture of positive- and negative-frequency functions
defined in another coordination, and the relation between them is given
by the Bogoliubov transformation. For example, a positive-frequency
function of Minkowski space charted by Rindler coordinates is a
mixture of positive- and negative-frequency functions of Minkowski space
charted by $ds^2=-dt^2+d{\bf x}^2$ (whose positive-frequency function
defines the Minkowski vacuum), and thus a comoving Rindler observer
(who experiences a 
constant acceleration in Minkowski space) detects particles (see
section~4.5 of ref.~\cite{birrell-davies}).

Sasaki, Tanaka, and Yamamoto  derive an appropriate
mixture of positive- and negative-frequency functions of de Sitter space
charted by open coordinates, whose annihilation operator annihilates the
Euclidean vacuum \cite{Sasaki:1994yt}. They expand $\hat\varphi$ as 
\begin{equation}
\hat{\varphi}(x)=\sum_{plm}\sum_{\sigma=\pm}\left[
u^{(\sigma)}_{plm}(x)\hat{a}^{(\sigma)}_{plm}+v^{(\sigma)}_{plm}(x)\hat{a}^{(\sigma)}_{plm}{}^\dagger\right]\,, 
\end{equation}
with $\hat{a}^{(\sigma)}_{plm}\left|0_E\right>=0$ for
$\sigma=\pm$. The mode functions, 
$u^{(\sigma)}_{plm}$ ($v^{(\sigma)}_{plm}$), are chosen to be
analytical in $E_2$ ($E_1$); hence the Euclidean vacuum. They are
linear combinations of positive- and negative-frequency
functions defined in the $R$ and $L$ regions.

They then write the mode functions as
$u^{(\sigma)}_{{plm}}(x)=u^{(\sigma)}_{p}(\eta)Y_{plm}({\bf x})$, where
$Y_{plm}({\bf x})$ are harmonics on a 3-hyperboloid, and the 
indices $p$, $l$, and  $m$ take on $0<p<\infty$, $l=0,1,2\cdots$, and
$m=-l,-l+1,\cdots,l-1,l$, respectively. The explicit form is given by
$Y_{plm}({\bf x})=f_{pl}(r)Y_{lm}(\Omega)$, where 
$Y_{lm}(\Omega)$ is the spherical harmonics on a 2-sphere, and
\cite{Sasaki:1994yt} 
\begin{align}
 f_{pl}(r) &=\left|\frac{\Gamma(ip+l+1)}{\Gamma(ip+1)}\right|
\frac{p}{\sqrt{\sinh{r}}} P^{-l-1/2}_{ip-1/2}(\cosh r)\,,
\label{eq:30}
\end{align}
with $P_\mu^\nu(z)$ being the associated Legendre functions.
This function goes as $f_{pl}(r)\propto e^{-r}$ for $p^2>0$, and
$r=1$ corresponds to the curvature radius of a 3-hyperboloid. Therefore,
the modes with $p^2>0$ decay exponentially on scales larger than the
curvature radius, and represent ``sub-curvature modes.'' In this
paper, we shall consider the modes with $p^2>0$ only, and ignore
``super-curvature modes''~\cite{Lyth:1995cw,GarciaBellido:1995wz}, which are described by the modes with
$p^2<0$. This is justified at the tree level: current observations
suggest that the curvature radius is greater than our current horizon
size, and the modes within our horizon are not affected by
super-horizon modes at the tree level.
The harmonics satisfy the following relations:
\begin{align}
\partial^2Y_{plm}({\bf x})&=-(p^2+1)Y_{plm}({\bf x})\,,\nnmb
Y^*_{plm}({\bf x})&=(-1)^mY_{{pl-m}}({\bf x})\,,\nnmb
\int d^3{\bf x}\sqrt{\gamma}\, Y^*_{{p_1l_1m_1}}({\bf x})Y_{p_2l_2m_2}({\bf x})
&=\delta(p_1-p_2)\delta_{l_1l_2}\delta_{m_1m_2}\,,\nnmb
\int_0^\infty dp\sum_{lm}Y^*_{plm}({\bf x})Y_{{plm}}({\bf
 x'})&=\delta^{(3)}({\bf x}- {\bf x}')\,. 
\label{eq:8}
\end{align}
The conformal-time dependence of the mode function
for a massless minimally-coupled scalar field,
$u_{p}^{(\sigma)}(\eta)$, is given by  \cite{Sasaki:1994yt}
\begin{equation}
u_{p}^{(\sigma)}(\eta)=
\frac{1}{\sqrt{2(1-e^{-2\pi p})}
}\tilde{u}_p(\eta)
-\sigma\frac{e^{-\pi p}}{\sqrt{2(1-e^{-2\pi p})}
}\tilde{v}_p(\eta)
\label{eq:mode1}
\end{equation}
with
\begin{align}
\tilde{u}_p(\eta)&= 
H_I\frac{\cosh\eta+ip\sinh\eta}{\sqrt{2p(1+p^2)}}
e^{-ip\eta},\qquad
\tilde{v}_p(\eta)= 
H_I\frac{\cosh\eta-ip\sinh\eta}{\sqrt{2p(1+p^2)}}
e^{ip\eta}\,.
\label{eq:37}
\end{align}
As $\tilde{u}_p$ and $\tilde{v}_p$ are positive- and
negative-frequency functions naturally defined in the $R$-region (recall
that eq.~\eqref{eq:37} is written in coordinates in the $R$-region), a
comoving observer in the $R$-region detects particles with respect to
the Euclidean vacuum. To see this, let us expand $\hat{\varphi}$ in the
$R$-region as $\hat{\varphi}(x)=\sum_{plm}[\tilde{u}_p(\eta)Y_{plm}({\bf
x})\tilde{a}_{plm}+\tilde{v}_p(\eta)Y^*_{plm}({\bf
x})\tilde{a}_{plm}^\dagger]$. Then, the occupation number of particles
is given by
\begin{equation}
 N_{plm}=\left<0_E\right|\tilde{a}^\dagger_{plm}\tilde{a}_{plm}\left|0_E\right>=
\frac1{e^{2\pi p}-1},
\end{equation}
which is a thermal spectrum. The argument and result given
here basically parallel those for particle creation in Rindler space,
i.e., a comoving observer in Rindler space sees particles with a thermal
spectrum (see eq.~(4.97) of ref.~\cite{birrell-davies}).

The Euclidean vacuum state given by eq.~\eqref{eq:mode1}
is a natural choice for quantum fluctuations in homogeneous de Sitter space
before bubble nucleation. The initial state then evolves away from the
Euclidean vacuum state via bubble nucleation; thus, eq.~\eqref{eq:mode1}
must be modified. 
In the $R$-region, we find $u_{\bf k}(x)$ by solving the following
linearized EOM, 
 \begin{align}
\frac{\delta L}{\delta\varphi}\equiv
 -\ddot{\varphi}(x)-
3H(t)\dot{\varphi}(x)+\frac{1}{a^2(t)}\partial^2\varphi(x)=0\,,
\label{eq:60}
\end{align}
where $\partial^2\equiv\gamma^{ij}\partial_i\partial_j$,
and $L$ is given by $S_2=\int d^4xa^3\sqrt{\gamma}~L$, and is also
related to the Lagrangian density, ${\cal L}$, by ${\cal
L}=a^3\sqrt{\gamma}L$. 
This EOM is obtained by the variation of the second order action given in
eq.~\eqref{eq:47}. 
For our model, we can choose the mode functions such that one vanishes
in the $R$-region and another vanishes in the $L$-region
\cite{Yamamoto:1996qq}. Specifically, we expand $\hat{\varphi}$ as
\begin{equation}
\hat{\varphi}(x)=\sum_{plm}\left[
u^{R}_{plm}(x)\hat{a}^{R}_{plm}+v^{R}_{plm}(x)\hat{a}^{R}_{plm}{}^\dagger
+u^{L}_{plm}(x)\hat{a}^{L}_{plm}+v^{L}_{plm}(x)\hat{a}^{L}_{plm}{}^\dagger
\right]\,,
\label{eq:varphiexpansion}
\end{equation}
where $u^{R}_{plm}$ and $u^{L}_{plm}$ vanish in the $L$-
and $R$-regions, respectively. (Similarly for $v_{plm}^R$ and
$v_{plm}^L$.) In this paper, we shall consider $u^R_{plm}$ (and
$v_{plm}^R$) only, as we have no access to the $L$-region 
observationally; 
henceforth we shall drop the superscript 
$R$, i.e., $u^R_{plm}\to u_{plm}$ and $v^R_{plm}\to v_{plm}$,
unless stated otherwise.
We then write
\begin{align}
u_{{plm}}(x)&=u_{p}(\eta)Y_{plm}({\bf x})\,.
\label{eq:35}
\end{align}
Yamamoto, Sasaki, and Tanaka find \cite{Yamamoto:1996qq}
\begin{equation}
u_{p}(\eta)=
\frac{1}{\sqrt{1-e^{-2\pi p}}
}\tilde{u}_p(\eta)
+\frac{e^{-\pi p-2i\delta_p}}{\sqrt{1-e^{-2\pi p}}
}\tilde{v}_p(\eta)\,,
\label{eq:mode2}
\end{equation}
with $\tilde{u}_p(\eta)$ and $\tilde{v}_p(\eta)$ given by eq.~\eqref{eq:37}.\footnote{Here, we do not give the explicit expression for the
complex phase, $\delta_p$, which does not affect the conclusion of this
paper (see ref.~\cite{Yamamoto:1996qq} for the way to obtain
$\delta_p$). While eq.~\eqref{eq:mode2} is written with the coordinates in the
$R$-region, analytical continuation lets us use it also in parts of the
$C$- and $E$-regions which are inside the bubble. }
Unlike the previous mode function given in eq.~\eqref{eq:mode1}, the
annihilation operator defined with respect to eq.~\eqref{eq:mode2}
does {\it not} annihilate the Euclidean vacuum. 
Again, $\tilde{u}_p$ and $\tilde{v}_p$ are positive- and
negative-frequency functions naturally defined in the $R$-region; thus,
writing
$\hat{\varphi}(x)=\sum_{plm}[\tilde{u}_p(\eta)Y_{plm}({\bf
x})\tilde{a}_{plm}+\tilde{v}_p(\eta)Y^*_{plm}({\bf
x})\tilde{a}_{plm}^\dagger]$ in the $R$-region, we find that the
occupation number of 
particles detected in the $R$-region with respect to the new 
vacuum state, $\left|0\right>$, is still given by a thermal spectrum,
$N_{plm}=\left<0\right|\tilde{a}^\dagger_{plm}\tilde{a}_{plm}\left|0\right>=
({e^{2\pi p}-1})^{-1}$.

As a comoving observer in
the $R$-region, which is inside the nucleated bubble, detects particles
(also see  
ref.~\cite{Yamamoto:1994te} which discusses particle creation due to
bubble nucleation in Minkowski space),
the state for quantum fluctuations set at the time of bubble nucleation
in open inflation is a {\it non}-Bunch-Davies vacuum state.

With the definition of the field operator given in
eq.~\eqref{eq:varphiexpansion}, 
the Wightman functions, $G^{\pm}(x,x')$, are given by
\begin{align}
 G^{+}(x,x')&\equiv\braket{0}{\varphi(x)\varphi(x')}{0}
=\sum_{plm}\left[u^R_{plm}(x)v^R_{plm}(x')+u^L_{plm}(x)v^L_{plm}(x')\right]\,,\nnmb 
 G^{-}(x,x')&\equiv\braket{0}{\varphi(x')\varphi(x)}{0}
=\sum_{plm}\left[u^R_{plm}(x')v^R_{plm}(x)+u^L_{plm}(x')v^L_{plm}(x)\right]\,. 
\label{eq:58}
\end{align}
While $G^{+}(x,x')=G^{-}(x,x')~(\equiv G(x,x'))$
for space-like separated $x$ and $x'$, $G^{+}(x,x')$ and $G^{-}(x,x')$
are different for time-like separated $x$ and $x'$. 

Let us now calculate the power spectrum. The multipole expansion
coefficients of $\varphi$ are defined by 
\begin{align}
 \varphi_{plm}(\eta)
&=\int  d^3{\bf x}\sqrt{\gamma}\,Y_{plm}^*({\bf x}) 
\varphi\left(\eta,{\bf x}\right)\,,
\label{eq:1}
\end{align}
and the power spectrum, $P(p)$, is defined by
$\big<\varphi^*_{plm}\varphi_{p'l'm'}\big>= 
\delta(p-p')\delta_{ll'}\delta_{mm'}P(p)$.
From eqs.~\eqref{eq:57}, \eqref{eq:8}, \eqref{eq:mode2}, and \eqref{eq:1},
$P(p)$ at late time (i.e., $\eta\approx 0$) is given by \cite{Yamamoto:1996qq}
\begin{align}
P(p)=H_I^2\frac{\cosh \pi p+\cos 2\delta_{p}}{2p(1+p^2)\sinh \pi p}
\label{eq:49}\,.
\end{align}
We recover a scale-invariant spectrum in the large $p$ limit,
$P(p)\to H_I^2/(2p^3)$. This is an expected result, as large-$p$ modes
do not feel curvature of space, and the occupation number of particles
falls off exponentially at large $p$.

%%%%%%%%%%%%%%%%%%%%%%%%%%%%%%%%%%%
%%%%%%%%%%%%%%%%%%%%%%%%%%%%%%%%%%%
\section{In-in formalism on a CDL instanton background}
\label{sec:qft}
%%%%%%%%%%%%%%%%%%%%%%%%%%%%%%%%%%%
%%%%%%%%%%%%%%%%%%%%%%%%%%%%%%%%%%%
Here, we outline the in-in formalism on a CDL instanton background,
which can be used to calculate the bispectrum in open inflation models.
The formalism is a natural extension of the in-in formalism in de Sitter
space charted by flat coordinates \cite{Maldacena:2002vr} to the case
of a universe with quantum tunneling. 
 The free QFT in open inflation, as reviewed in
section~\ref{sec:mode_open}, is based on a WKB analysis of a tunneling wave
function without non-linear interaction
terms~\cite{Yamamoto:1993mp,Tanaka:1993ez,Tanaka:1994qa}.  By 
performing a similar analysis with interaction terms, 
the in-in formalism can be extended to the case of 
open inflation~\cite{Sugimura:2012kr,Sugimura:2013sea,Park:2011ty}.

\begin{figure}
\begin{center}
 \includegraphics[width=10cm]{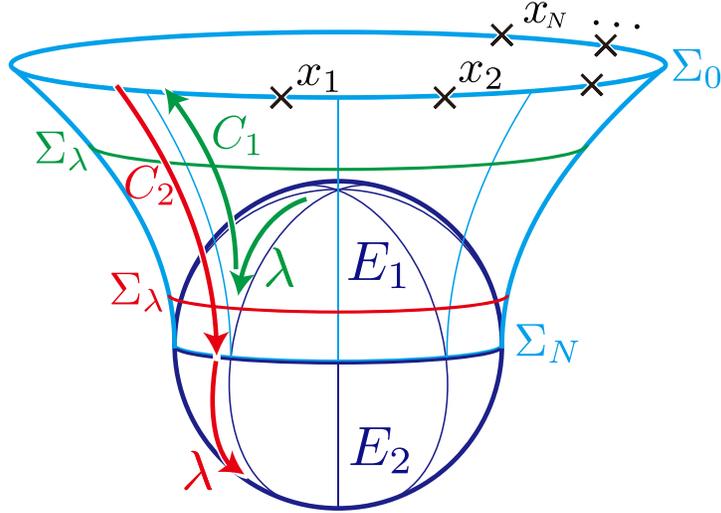} 
\end{center}
\caption{
A 4-dimensional CDL instanton embedded in 5-dimensional Minkowski
spacetime. The integral that gives an $N$-point correlation 
 function on a spatial hypersurface $\Sigma_0$ (eq.~\eqref{eq:24}) is
 performed along a path, $\lambda$, and 
 over space, ${\bf x}$. The $\lambda$ integration is done along
 $C=C_1+C_2$, and the ${\bf x}$ integration is done over a
 hypersurface $\Sigma_\lambda$ for a given $\lambda$.  The upper
 cylinder (light blue) is the 
Lorentzian region after bubble nucleation.  The upper and lower
hemispheres (dark blue) are the Euclidean regions $E_1$ and $E_2$,
 respectively. A bubble is assumed to be nucleated on the hypersurface
$\Sigma_N$. (The bubble wall is not shown.)  The hypersurfaces
 $\Sigma_\lambda$ are chosen 
 to be Cauchy surfaces in the 
Lorentzian region, and are chosen to cover $E_1$ and $E_2$ in the Euclidean
regions of the integration domain $V_1$ (the green arrows ending on
 $\Sigma_0$) and $V_2$ (the red arrows ending in $E_2$), respectively.}
\label{fig:inin_path}
\end{figure}

The Lagrangian density is given by
\begin{align}
 \mathcal{L}_{full}&=\mathcal{L}_0+\mathcal{L}_{int}\,,
\end{align}
where $\mathcal{L}_0$ and $\mathcal{L}_{int}$ are the free and interaction
parts of the Lagrangian density, respectively.
Non-trivial time evolution in $\mathcal{L}_0$ causes the state to evolve
away from 
the Bunch-Davies vacuum state,
and $\mathcal{L}_{int}$ is a source of non-Gaussianity. 
The in-in formalism on a CDL instanton background tells us that 
the $N$-point function of $\varphi$ for $x_i$s on a spacial
hypersurface $\Sigma_0$ is given by  
\begin{align}
\Big<\varphi(x_1)\varphi(x_2)\cdots\varphi(x_N)\Big>
=& \frac{\braket{0}
{P\,\varphi(x_1)\varphi(x_2)\cdots\varphi(x_N)
e^{i\int_{C\times \Sigma_{\lambda }}d\lambda d^3{\bf x}\,\mathcal{L}_{int}(x)}}{0}}
{\braket{0}
{P
e^{i\int_{C\times \Sigma_{\lambda }}d\lambda d^3{\bf x}\,\mathcal{L}_{int}(x)}
}{0}}\,,
\label{eq:24}
\end{align}
where the $\lambda$ integral is performed along the path $C=C_1+C_2$,
and the ${\bf x}$ integral is over the hypersurface $\Sigma_\lambda$
for a given $\lambda$ (see figure~\ref{fig:inin_path}).
The path ordering operator, $P$, orders operators in the expression
according to the order along $C$. 
Covariance of eq.~\eqref{eq:24} guarantees that
the $N$-point function is not affected by the choice of $C$ or $\Sigma_\lambda$.

The path of integration is shown in figure~\ref{fig:inin_path}.
In the first integration domain, $V_1\equiv C_1\times\Sigma_{\lambda}$
(the green arrows), the path $C_1$ starts in $E_1$ and runs through the
nucleation surface 
$\Sigma_N$ to $\Sigma_0$. In the second domain, $V_2\equiv
C_2\times\Sigma_{\lambda}$ (the red arrows), the path $C_2$ starts from
$\Sigma_0$ and runs through $\Sigma_N$ into $E_2$. 
The contributions to the integral from the Euclidean and Lorentzian regions
correspond to contributions from during and after bubble nucleation,
respectively. 

To evaluate eq.~\eqref{eq:24}, we use Wick's theorem
after Taylor expanding the exponential.
Due to the path-ordering operator, $P$,
each pair of field operators becomes $G_{P}(x,x')$,
which is given by
\begin{align}
 G_{P}(x,x')&\equiv
\braket{0}{P\varphi(x)\varphi(x')}{0}
=
\begin{cases}
 G^+(x,x')&{\rm when\ }x'{\rm\ precedes\ } x\\
 G^-(x,x')&{\rm when\ }x{\rm\ precedes\ } x'
\end{cases}\,,
\label{eq:68}
\end{align}
where $G^{\pm}(x,x')$ are the Wightman functions given by eq.~\eqref{eq:58}.

Let us briefly comment on analyticity of $G_{P}(x,x')$
(see appendix~\ref{sec:tunneling-region} for details).
This function is singular when $x$ and $x'$ are null-separated.
However, by assuming a small imaginary part in the time coordinates along 
the in-in path in eq.~\eqref{eq:24}, $G_{P}(x,x')$ becomes analytical.
As a result, $\big<\varphi(x_1)\varphi(x_2)\cdots\varphi(x_N)\big>$ is
also analytical with respect to $x_i$s.

%%%%%%%%%%%%%%%%%%%%%%%%%%%%%%%%%%%
\section{Bispectrum from open inflation}
\label{sec:ng-gen}
\subsection{Cubic action in the flat gauge}
%%%%%%%%%%%%%%%%%%%%%%%%%%%%%%%%%%%
To calculate the bispectrum at the tree-level, we substitute the
constraint equations into the original action, keep $O(\varphi^3)$
quantities, and obtain  the reduced
third-order action in the flat gauge (in which 
$\zeta\equiv 0$).

The third-order action to the leading order in the slow-roll parameters is
$O(\epsilon^{1/2})$ and given by (see appendix~\ref{sec:int_by} for derivation)
\begin{align}
 S_3&=
-\int dtd^3{\bf x}\sqrt{\gamma}~a^5\dot\phi\left[(\partial^2-3)^{-1}\dot{\varphi}\right]
\dot{\varphi}^2\nnmb
&+
\int dtd^3{\bf x}\sqrt{\gamma}\left\{
a^5\ddot{\phi}\left[(\partial^2-3)^{-1}\varphi\right]\dot{\varphi}^2
+\frac{3a^3\dot{\phi}}{4H}\left[(\partial^2-3)^{-1}\varphi\right]
\left(\dot{\varphi}^2
+\frac{1}{a^2}\partial_i\varphi\partial^i\varphi\right)\right\}\nnmb
&
+\int dtd^3{\bf x}\left\{-\frac{a^2\dot{\phi}}{4H}
\left[(\partial^2-3)^{-1}
\left(\dot{\varphi}^2+\frac{1}{a^2}\partial_i\varphi\partial^i\varphi \right)\right]
-\frac{a^2}{2H}\left[(\partial^2-3)^{-1}\dot{\varphi}\right]
\left(\dot{\phi}\dot{\varphi}
-\ddot{\phi}\varphi\right)\right\}
\frac{\delta {\cal L}}{\delta \varphi}\,,
\label{eq:29}
\end{align}
where $\delta{\cal L}/\delta\varphi\equiv a^3\sqrt{\gamma}\delta
L/\delta\varphi$,  with $\delta L/\delta \varphi$ given in eq.~\eqref{eq:60},
and $\partial^2\equiv
\gamma^{ij}\partial_i\partial_j$. 
Let us check the correspondence between this action and that found by
Clunan and Seery \cite{Clunan:2009ib}, who give the third-order action with
positive spatial curvature. 
\begin{itemize}
\item We need to replace their $\partial^2+3$
(denoted as $\Delta+3$) with $\partial^2-3$, as we deal with
negative curvature. 
\item The first line in eq.~\eqref{eq:29} agrees with the second term in
      their eq.~(4.5).
\item The second line in eq.~\eqref{eq:29} does not appear in their action:
the first term proportional to $\ddot{\phi}$ is ignored in their action
because it is a higher order in slow-roll in their model. While
      $\ddot{\phi}/(H\dot{\phi})$ is slow-roll suppressed in the usual 
inflation scenario, it is not so in open inflation because
$\ddot{\phi}/(H\dot{\phi})={\cal O}(1)$ soon after bubble nucleation.
The second
term proportional 
to $3a^3\dot\phi/(4H)$ does not appear in their action due to the
difference in EOM. Had our EOM had an extra mass term like theirs, the
second term would cancel out with a term in the second line.
\item The last line of eq.~\eqref{eq:29} agrees with the last term in
their eq.~(4.5) with $f(\varphi)$ given in their eq.~(4.6), except
for the sign of the second term in 
their eq.~(4.6). This is because their second-order action (eq.~(2.15))
contains an extra mass term of $V''=-3/a^2$. 
\item The first term in their eq.~(4.5), $(aH/\dot\phi)\varphi^3$, which
      comes from $V'''$, is ignored in our action.
\end{itemize}

As in the case of the second order action,
we expect that this expression derived in the $R$-region
can also be used outside the $R$-region by analytical continuation,
except near the bubble wall, for which a non-zero mass term should be
inserted into the action. Potentially large self-interaction terms near
the bubble wall do not contribute to the bispectrum in the sub-curvature
approximation, as discussed in appendix~\ref{sec:tunneling-eval-each}.

The terms in the second line are proportional to 
the EOM; thus, we remove it by making the field redefinition
\cite{Maldacena:2002vr},  
$\varphi\to\varphi_{c}$, where 
 \begin{align}
 \varphi&=\varphi_c+
 \frac{\dot{\phi}}{4H}\left[(\partial^2-3)^{-1}\partial_i\varphi_c\partial^i\varphi_c\right]
 +\cdots\,,
\label{eq:26}
 \end{align}
where $\cdots$ contains terms which vanish outside the horizon.
The field redefinition does not affect the second-order action, $S_2$,
given in eq.~\eqref{eq:47}. The positive-frequency function for
$\varphi_{c}$ is given by eqs.~\eqref{eq:35} with \eqref{eq:mode2}.

We write the interaction Lagrangian density, $\mathcal{L}_{int}$,
in terms of  $\varphi_{c}$ as
 \begin{align}
 \mathcal{L}_{int}=
\sqrt{\gamma}&\left\{
-a^5\dot{\phi}\left[(\partial^2-3)^{-1}\dot{\varphi_{c}}\right]\dot{\varphi_{c}}^2 
+a^5\ddot{\phi}\left[(\partial^2-3)^{-1}\varphi_{c}\right]\dot{\varphi_{c}}^2\right.\nnmb
&\left.+\frac{3a^3\dot{\phi}}{4H}\left[(\partial^2-3)^{-1}\varphi_{c}\right] 
\left(\dot{\varphi_{c}}^2
+\frac{1}{a^2}\partial_i\varphi_{c}\partial^i\varphi_{c}\right)\right\}\,.
\label{eq:39}
 \end{align}
If we write eq.~\eqref{eq:26} schematically as
$\varphi=\varphi_c+\lambda\varphi_c^2$, the 3-point function of
$\varphi$ is given by 
\begin{align}
\Big<\varphi(x_1)\varphi(x_2)\varphi(x_3)\Big>
=&\Big<\varphi_c(x_1)\varphi_c(x_2)\varphi_c(x_3)\Big>\nnmb
&+\lambda\left[\Big<\varphi_c(x_1)\varphi_c(x_2)\Big>
\Big<\varphi_c(x_1)\varphi_c(x_3)\Big>+{\rm (perms.)}\right]\,,
\label{eq:11a}
\end{align}
where $\big<\varphi_c(x_1)\varphi_c(x_2)\varphi_c(x_3)\big>$ is obtained
by substituting $\mathcal{L}_{int}$ (eq.~\eqref{eq:39})
into the expression for the $N$-point function given in eq.~\eqref{eq:24}.

%%%%%%%%%%%%%%%%%%%%%%%%%%%%%%%%%%%
\subsection{An example calculation}
%%%%%%%%%%%%%%%%%%%%%%%%%%%%%%%%%%%
Before we compute the bispectrum using the cubic action given by
eq.~\eqref{eq:39}, let us show how the calculation proceeds using a
simpler example. Consider $\mathcal{L}_{int}$ given by 
\begin{align}
 \mathcal{L}_{int}(x)=\sqrt{-g}\,\lambda_{int}(t)\varphi^3(x)\,.
\label{eq:69}
\end{align}
Although $\mathcal{L}_{int}(x)$ that we wish to use
is not in this form, both  eqs.~\eqref{eq:39} and \eqref{eq:69} respect
O(4)-symmetry of the background universe; thus, the calculations
proceed essentially in the same way.

Using eq.~\eqref{eq:24}, Wick's theorem, and eq.~\eqref{eq:68}, we obtain
the tree-level 3-point function for space-like separated points
$x_1$, $x_2$, and $x_3$ as
\begin{align}
\Big<\varphi(x_1)\varphi(x_2)\varphi(x_3)\Big>
=&2\re\left[-i\int_{V_2}d^4x
 \sqrt{-g}\,\lambda_{int}(t)G^{+}(x,x_1)G^{+}(x,x_2)G^{+}(x,x_3)\right]
+{\rm (perms.)}\,,
\label{eq:63}
\end{align}
As the direct evaluation of the real-space expression such as eq.~\eqref{eq:63} 
is often not practical, we shall move to the harmonic-space expression by
operating $\prod_{i=1}^3\int d^3{\bf x}_i\sqrt{\gamma}\,Y^*_{p_il_im_i}({\bf x}_i)$
on both sides of eq.~\eqref{eq:63}, and using eqs.~\eqref{eq:58},
\eqref{eq:35}, \eqref{eq:8}, and \eqref{eq:1}.
We shall only consider the modes with $p^2>0$. Orthogonality of
$Y_{p_il_im_i}$ then implies that super-curvature modes with $p^2<0$ do
not contribute.

As a result of O(4)-symmetry of the system, the harmonic-space
expression always contains a geometrical factor given by 
 $\int d^3{\bf x}\sqrt{\gamma}\,Y_{p_1l_1m_1}({\bf x})Y_{p_2l_2m_2}({\bf x})
Y_{p_3l_3m_3}({\bf x})
=\mathcal{F}_{p_1p_2p_3}^{l_1l_2l_3}\mathcal{G}^{m_1m_2m_3}_{l_1l_2l_3}\,$,
where 
\begin{align}
 \mathcal{F}_{p_1p_2p_3}^{l_1l_2l_3}
&\equiv
\int dr~ \sinh^2 r 
f_{p_1l_1}(r)f_{p_2l_2}(r)f_{p_3l_3}(r)\,,
\label{eq:51}\\
\mathcal{G}^{m_1m_2m_3}_{l_1l_2l_3}
&\equiv
\int d^2\Omega~ Y_{l_1m_1}(\Omega)Y_{l_2m_2}(\Omega)Y_{l_3m_3}(\Omega)\,.
\label{eq:51a}
\end{align}
See appendix~\ref{sec:tunneling-region} for the issue on
the integration outside the $R$-region in $V_2$ in eq.~\eqref{eq:63}.
The bispectrum, $B(p_1,p_2,p_3)$, is given by
\begin{equation}
\big<\varphi_{p_1l_1m_1}\varphi_{p_2l_2m_2}\varphi_{p_3l_3m_3}\big>
= B(p_1,p_2,p_3)
\mathcal{F}_{p_1p_2p_3}^{l_1l_2l_3}\mathcal{G}^{m_1m_2m_3}_{l_1l_2l_3}.
\end{equation}
As we show in appendix~\ref{sec:tunneling-eval-each}, the contribution
to the bispectrum in the sub-curvature approximation ($p\gg 1$) mainly
comes from the $R$-region, and the bispectrum at late time
(i.e., $\eta\approx 0$) is given by 
\begin{align}
 B(p_1,p_2,p_3)\approx 2\re&\left\{-iv_{p_1}(0)v_{p_2}(0)v_{p_3}(0)
  \left[\int_{-\infty}^{0} a^4(\eta) d\eta\,\lambda_{int}(\eta)
u_{p_1}(\eta)u_{p_2}(\eta)u_{p_3}(\eta)
\right]\right.\nnmb
&+{\rm (perms.)} 
\Big\}\,.
\label{eq:15}
\end{align}
The contribution from outside the $R$-region can be included by
extending the integration domain to the complex plane 
(see eq.~\eqref{eq:48} in appendix~\ref{sec:tunneling-region}). 
Whilst eq.~\eqref{eq:15} was obtained from $\mathcal{L}_{int}$ given in
eq.~\eqref{eq:69}, $B(p_1,p_2,p_3)$ calculated from 
the true $\mathcal{L}_{int}$ given in eq.~\eqref{eq:39}
takes a similar form. 

%%%%%%%%%%%%%%%%%%%%%%%%%%%%%%%%%%%%%%%%%%%%%%%
 \subsection{Bispectrum in the sub-curvature approximation}
\label{sec:sq_ng}
%%%%%%%%%%%%%%%%%%%%%%%%%%%%%%%%%%%%%%%%%%%%%%%
In this section we calculate the bispectrum from open inflation in
the sub-curvature approximation, $p\gg 1$. This approximation is
justified because we have not yet detected spatial curvature within our
current horizon. As we show in appendix~\ref{sec:sque-limit-mode},
the power spectrum, $P(p)$, and the bispectrum, $B(p_1,p_2,p_3)$, reduce
to $P(k)$ and $B(k_1,k_2,k_3)$ in the sub-curvature
approximation with $p_i\to k_i$,
where $k_i$'s denote Fourier wavenumbers defined in flat space.
Therefore, we shall use $k_i$ for the indices of harmonics instead of $p_i$.

We calculate the bispectrum using the same method as was used to derive
eq.~\eqref{eq:15}, but replacing $\mathcal{L}_{int}$ in eq.~\eqref{eq:69} with that given in eq.~\eqref{eq:39}.

%%%%%%%%%%%%%%%%%%%%%%%%%%%%%%%%%%%%%%%%%%%%%%%
\subsubsection{Bispectrum of $\varphi$}
%%%%%%%%%%%%%%%%%%%%%%%%%%%%%%%%%%%%%%%%%%%%%%%
First, let us compute the bispectrum from the field redefinition
term. Recalling $P(k)=H_I^2/(2k^3)$ in the sub-curvature approximation,
we find
\begin{align}
B^{(\rm redef)}(k_1,k_2,k_3)
&=\frac{\dot{\phi}}{4H_I}\frac{\bf{k}_1\cdot\bf{k}_2}{k_3^2+4}\frac{H_I^2}{2k_1^3}\frac{H_I^2}{2k_2^3}+{\rm
(perms.)}\,.
\label{eq:bispecredef}
\end{align}

Second, we compute the bispectrum from the in-in integral.
As we show in appendix~\ref{sec:tunneling-eval-each}, the
leading-order term in the sub-curvature approximation is the first term in
eq.~\eqref{eq:39},
\begin{align}
 \mathcal{L}^{(1)}_{int}&\equiv
-\sqrt{\gamma}
a^5\dot{\phi}\left[(\partial^2-3)^{-1}\dot{\varphi_{c}}\right]\dot{\varphi_{c}}^2\,, 
\label{eq:25}
\end{align}
and the other terms are sub-dominant. We thus compute
\begin{align}
 B(k_1,k_2,k_3)
 =
 2\re\left\{iv_{k_1}(0)v_{k_2}(0)v_{k_3}(0)
 \left[\int_{-\infty}^0 d\eta  
 \frac{a^6\dot{\phi}}{k_1^2+4}
\dot{u}_{k_1}(\eta)\dot{u}_{k_2}(\eta)\dot{u}_{k_3}(\eta)\right]
+{\rm (perms.)} 
\right\}\,.
\label{eq:21}
\end{align}

The mode functions, $u_k(\eta)$, $\dot u_k(\eta)$, and  $v_k(0)$,
are given by (see eq.~\eqref{eq:mode2})
\begin{align}
 u_k(\eta)&\approx 
\begin{cases}
\ds -\frac{i}
{a(\eta)\sqrt{2k}}
\left(e^{-ik\eta}-e^{-\pi k-2i\delta_k}e^{ik\eta}
\right)&\left(-\infty<\eta\lesssim -1\right)\\[10pt]
\ds\frac{H_I}
{\sqrt{2k^3}}\left[(1+ik\eta)e^{-ik\eta}-
e^{-\pi k-2i\delta_k}(1-ik\eta)e^{ik\eta}\right]
&\left(-1 \lesssim \eta< 0\right)
\end{cases}
\,,\label{eq:31}\\
 \dot{u}_k(\eta)&\approx\frac{1}{a^2(\eta)}\sqrt{\frac{k}{2}}
\left(e^{-ik\eta}
+e^{-\pi k-2i\delta_k}e^{ik\eta}\right)\,,\label{eq:44}\\
v_k(0)&=u^*_k(0)\approx \frac{H_I}{\sqrt{2k^3}}(1-e^{-\pi k-2i\delta_k})\,,
\end{align}
where we have kept the terms proportional to $e^{-\pi k}e^{ik\eta}$,
despite that they are suppressed by $e^{-\pi k}$. This is because these
terms represent the effect of a non-Bunch-Davies initial state, and we shall
discuss the effect of these terms toward the end of this subsection.

\begin{figure}
\begin{center}
 \includegraphics[width=10cm]{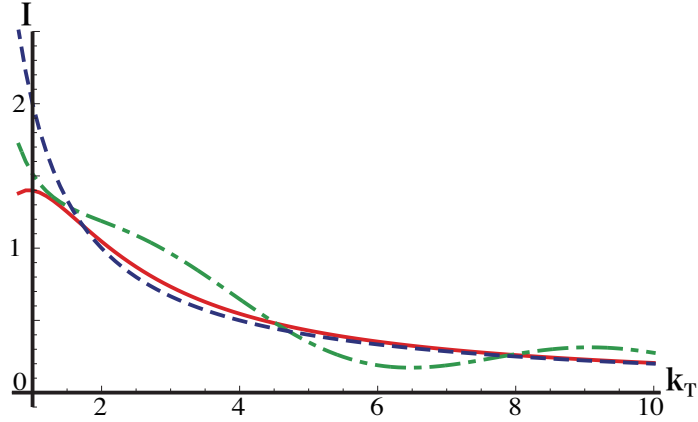} 
\end{center}
\caption{
We calculate the integral given by $I\equiv
 (1/2\sqrt{\epsilon}H_I)2\re[i\int_{-\infty}^{0}d\eta~ \dot{\phi} 
e^{-ik_T\eta} ]$ with $k_T\equiv k_1+k_2+k_3$ in three
 ways. First, we use the numerical solution of eq.~\eqref{eq:67} for
 $\dot\phi$  in the integral [solid (red) line]. Second, 
we use the approximate solution for $\dot\phi$ given in ~\eqref{eq:36}
 [dot-dashed (green) line]. This gives the term in the square
 bracket in eq.~\eqref{eq:79} divided by $k_T$.
Finally, we remove the oscillatory term from the second case [dashed
 (blue) line]. This result is in good agreement with the full solution
 (solid line) for sub-curvature modes ($1\ll k_T$), justifying our
 ignoring the oscillatory term in eq.~\eqref{eq:40} and thereafter.
}
\label{fig:num_int}
\end{figure}

Ignoring the non-Bunch-Davies terms for the moment, we obtain
\begin{align}
 B(k_1,k_2,k_3)
=
2\re\left[i\left(2\sum_{i=1}^3\frac{H_I^3}{k_i^2}\right)
\frac{1}{8k_1k_2k_3}
\int_{-\infty}^{0}d\eta~ \dot{\phi}
e^{-i(k_1+k_2+k_3)\eta} \right]\,.
\label{eq:38}
\end{align}
Using eq.~\eqref{eq:36} for $\dot\phi$, the integration during the
curvature-dominated era (i.e., $-\infty<\eta<-1$) is given by 
 \begin{align}
 \int_{-\infty}^{-1}d\eta~ \dot{\phi}
 e^{-i(k_1+k_2+k_3)\eta}
 =
 -\frac{3\sqrt{2\epsilon}H_I}{2}\frac{
 e^{-1+i(k_1+k_2+k_3)}}
 {1-i(k_1+k_2+k_3)}\,,
\label{eq:52}
 \end{align}
and that during the inflationary era (i.e., $-1<\eta<0$) is given by 
 \begin{align}
 \int_{-1}^{0}d\eta~ \dot{\phi}
 e^{-i(k_1+k_2+k_3)\eta}
 =
 -\sqrt{2\epsilon}H_I
 \frac{1-e^{i(k_1+k_2+k_3)}}
 {-i(k_1+k_2+k_3)}\,.
\label{eq:53}
 \end{align}
We thus find, in $k\gg 1$, 
\begin{align}
 B(k_1,k_2,k_3)
=
\frac{\sqrt{2\epsilon}H_I^4}{4k_1k_2k_3(k_1+k_2+k_3)}
\left(\sum_{i=1}^{3}\frac{1}{k_i^2}\right)
\Big[2-0.9\cos(k_1+k_2+k_3)\Big]\,,
\label{eq:79}
\end{align}
where the factor $0.9$ comes from $-2+3e^{-1}\approx-0.9$.
This result is based upon the approximate form for $\dot\phi$ given in
eq.~\eqref{eq:36}, which is discontinuous at $\eta=-1$. We find that
the second oscillatory term in eq.~\eqref{eq:79} is an artifact due to
this discontinuity.\footnote{We would like to thank T. Tanaka for
pointing this out.} To show this, we use the numerical solution of
eq.~\eqref{eq:67} for $\dot\phi$ and evaluate the integral. In
figure~\ref{fig:num_int}, we show that the numerical solution does not
contain any oscillations. Henceforth we shall neglect the oscillatory
terms coming from the discontinuity of the approximate solution of
$\dot\phi$. We obtain 
\begin{align}
 B(k_1,k_2,k_3)
=
\frac{2\sqrt{2\epsilon}H_I^4}{4k_1k_2k_3(k_1+k_2+k_3)}
\left(\sum_{i=1}^{3}\frac{1}{k_i^2}\right)
\,,
\label{eq:40}
\end{align}
which agrees with the result for the single-field slow-roll inflation
models with the canonical kinetic term \cite{Maldacena:2002vr}. 
In the squeezed limit, $k_3\ll k_1\approx
k_2\equiv k$, we find
\begin{equation}
B(k_1,k_2,k_3) \to 
\frac{\sqrt{2\epsilon}H_I^4}{4k^3k_3^3}=
\sqrt{2\epsilon}P(k)P(k_3)\,.
\label{eq:squeezed}
\end{equation}
The bispectrum from the field redefinition term 
(eq.~\eqref{eq:bispecredef}) in the squeezed limit goes as 
\begin{equation}
B^{(\rm redef)}(k_1,k_2,k_3)\to
-\frac{\sqrt{2\epsilon}H_I^4}{16}\frac1{k^4k_3^2}
=
-\sqrt{2\epsilon}\frac{k_3}{4k}P(k)P(k_3)\,.
\end{equation}
Thus, the field redefinition term is sub-dominant in this limit.

%%%%%%
Next, let us discuss an enhancement of the bispectrum 
due to a non-Bunch-Davies vacuum initial state, $B^{({\rm
NBD})}(k_1,k_2,k_3)$. 
The relevant term in eq.~\eqref{eq:44} is the
part of $\dot{u}_{k}(\eta)$ that is proportional to $e^{ik\eta}$ (i.e.,
$\dot{\tilde{v}}_{k}(\eta)$) instead of $e^{-ik\eta}$ (i.e.,
$\dot{\tilde{u}}_{k}(\eta)$).  We then do this flipping in each of
$\dot{u}_{k_i}$ in the integral of eq.~\eqref{eq:21}. For example,
flipping $\dot{u}_{k_1}$ we obtain
\begin{align}
& B^{( \tilde{v}_{k_1})}(k_1,k_2,k_3) \nnmb
&=
 2\re\left[i\left(2\sum_{i=1}^3\frac{H_I^3}{k_i^2}\right)
\frac{e^{-\pi k_1-2i\delta_{k_1}}}{4k_1k_2k_3}
\int_{-\infty}^{0}d\eta~ \dot{\phi}
e^{-i(-k_1+k_2+k_3)\eta}
\right]\nnmb
&=
 \frac{\sqrt{2\epsilon}H_I^4e^{-\pi k_1}}{8k_1k_2k_3}
\left(\sum_{i=1}^{3}\frac{1}{k_i^2}\right)
\re\left[e^{-2i\delta_{k_1}}\left(
2
\frac{1-e^{i(-{k_1}+ {k_2}+ {k_3})}}
{-k_1+ k_2+ k_3}
-3i
\frac{e^{-1+i(-{k_1}+ {k_2}+ {k_3})}}
{1-i(-k_1+ k_2+ k_3)}\right)
\right]\,.
\label{eq:54}
\end{align}
Flipping $\dot{u}_{k_2}$ or $\dot{u}_{k_3}$ will similarly give
$B^{( \tilde{v}_{k_2})}$ or $B^{(\tilde{v}_{k_3})}$; hence $B^{(\rm
NBD)}=\sum_{i=1}^3B^{(\tilde{v}_{k_i})}$ to the leading order of
$e^{-\pi k}$.
We are interested in a peculiar behavior of the
bispectrum from a non-Bunch-Davies initial state in the squeezed limit,
$k_3\ll k_1\approx k_2\equiv k$
\cite{Agullo:2010ws,Ganc:2011dy,Chialva:2011hc}, 
while still satisfying the sub-curvature approximation, $1\ll
|-k_1+k_2+k_3|$. Again ignoring the oscillatory term which
is an artifact of using the approximate form of $\dot{\phi}$ in
eq.~\eqref{eq:36}, we find
\begin{align}
 B^{({\rm NBD})}(k_1,k_2,k_3)&\to 2\sqrt{\frac{\epsilon}{2}}H_I^4\,
\cos(2\delta_k)
\frac{e^{-\pi k}}{ k_3^{4}k^{2}}\nnmb
&= 4\sqrt{2\epsilon}\,
\cos(2\delta_k)
e^{-\pi k}\frac{k}{k_3}P(k)P(k_3)\,.
\label{eq:28}
\end{align}
There is an enhancement factor in the squeezed limit, $k/k_3\gg 1$,
compared to the usual single-field model with a Bunch-Davies initial
state, $B(k,k,k_3)\propto k_3^{-3}k^3$ (see eq.~\eqref{eq:squeezed}).
This result is in agreement with the previous work on the effect of
modified initial states on the bispectrum, in which modifications are given
by hand \cite{Agullo:2010ws,Ganc:2011dy,Chialva:2011hc}.
Despite the enhancement factor by $k/k_3$, however, $B^{({\rm
NBD})}(k_1,k_2,k_3)$ is always subdominant in the sub-curvature
approximation due to the exponential suppression factor, $e^{-\pi k}$. 

Let us now consider the exact folded limit, $k_i=k_j+k_l$. In the
previous study on the effect of modified initial states (given by hand) on the
bispectrum, the bispectrum diverges in this limit
\cite{Chen:2006nt,Holman:2007na,Meerburg:2009ys}. However, a
self-consistent computation presented here gives a finite result.
There is one subtlety: in order for us to use Fourier wavenumbers
defined in flat space, $k_i$, we need to make sure that $k_i+k_j-k_l$
for any combinations of $i$, $j$, and $k$ are greater than unity (i.e.,
sub-curvature approximation). This requirement is not met in the exact
folded limit. We thus go back to the open harmonics with $p_i$.
Also taking the squeezed limit, $p_3\ll p_1\approx p_2\equiv p$, for simplicity,
we obtain a compact formula, 
\begin{align}
 B^{({\rm NBD})}(p,p-p_3,p_3)&\to -3.1\sqrt{\frac{\epsilon}{2}}H_I^4\,
\sin(2\delta_p)
\frac{e^{-\pi p}}{ p_3^{3}p^{2}} \,,
\label{eq:61}
\end{align}
which remains finite, and is suppressed by $e^{-\pi p}$.

%%%%%%%%%%%%%%%%%%%%%%%%%%%%%%%%%%%%%%%%%%%%%%%
\subsubsection{Bispectrum of $\zeta$}
%%%%%%%%%%%%%%%%%%%%%%%%%%%%%%%%%%%%%%%%%%%%%%%
To compute the late-time observables such as the temperature and
polarization 
anisotropy of the CMB, we need to compute the curvature perturbation on
a uniform-density hypersurface, $\zeta$, which is equal to a
``comoving'' ($\varphi=0$) hypersurface and is conserved outside the
horizon. We thus perform the following non-linear transformation relating the
scalar-field fluctuation in the flat gauge, $\varphi$, to $\zeta$
\cite{Salopek:1990jq} (also see eq.~(2.66) of ref.~\cite{Komatsu:2002db})
\begin{align}
\zeta &= \int_{\phi_*+\varphi_*}^{\phi_*}d\phi~\frac{H}{\dot\phi}\nnmb
&=-\frac{H_*}{\dot{\phi}_*}\varphi_*
-\left.\frac{1}{2}\frac{\partial}{\partial\phi}\left(\frac{H}{\dot{\phi}}\right)\right|_{\phi=\phi_*}\varphi_*^2+{\cal O}(\varphi_*^3)\nnmb
&=-\frac{H_*}{\dot{\phi}_*}\varphi_*
+\frac{1}{2}\left(\frac{H_*\ddot{\phi}_*}{\dot{\phi}_*^3}
+\frac{1}{2}\right)\varphi_*^2+{\cal O}(\varphi_*^3)\,,
\label{eq:55}
\end{align}
where the subscript $*$ indicates the moment of the horizon exit, and
$H_*\approx H_I$ during slow-roll inflation.
The first term in eq.~\eqref{eq:55} implies that the bispectrum of
$\varphi$ computed in the previous section times $-H_*^3/\dot{\phi}_*^3$
 yields the bispectrum of $\zeta$. In addition, the second-term yields
 an additional contribution to the bispectrum of $\zeta$.

In the sub-curvature approximation, the power spectrum of $\varphi$ given in
eq.~\eqref{eq:49} multiplied by
$H_*^2/\dot\phi_*^2=1/(2\epsilon_*)$ with $p\to k$  yields
\begin{equation}
 P_\zeta(k) = \frac{H_*^2}{4\epsilon_*}\frac1{k^3}.
\end{equation}
Using the relation $k=a_*H_*$ and taking into account the
slow-roll time-dependence of $H_*$ and $\epsilon_*$, we obtain the power
spectrum index as
\begin{equation}
 n_s-1\equiv \frac{d\ln k^3P_\zeta(k)}{d\ln k} = 
-2\frac{\ddot{\phi}_*}{H_*\dot{\phi}_*}-2\frac{\dot{\phi}_*^2}{H_*^2}.
\end{equation}

Ignoring the non-Bunch-Davies terms and taking the squeezed limit,
$k_3\ll k_1\approx k_2\equiv k$, we obtain 
from eqs.~\eqref{eq:squeezed} and \eqref{eq:55}
\begin{align}
 B_\zeta(k_1,k_2,k_3)
&\to
-\frac{H_I^3}{\dot{\phi}_*^3}\frac{H_I^4\sqrt{2\epsilon_*}}{4k^3k_3^3}
+\frac{2H_I^2}{\dot{\phi}_*^2}
\left(\frac{H_I\ddot{\phi}_*}{\dot{\phi}_*^3}+\frac{1}{2}\right)
\frac{H_I^2}{2k_3^3}\frac{H_I^2}{2k^3}
\nnmb
&=
(1-n_{s})
P_\zeta(k)P_\zeta(k_3)\,,
\label{eq:59}
\end{align}
where we have used $\sqrt{2\epsilon_*}=-\dot\phi_*/H_I$ (recall
$\dot\phi<0$ during slow-roll inflation in our model).
This result agrees with the squeezed limit of the usual single-field
inflation models with a Bunch-Davies initial state  \cite{Maldacena:2002vr}. 
Therefore, aside from the terms suppressed in the sub-curvature
approximation, Maldacena's 
consistency relation, $B_\zeta\to (1-n_s)P_\zeta(k)P_\zeta(k_3)$ as
$k_3\to 0$, holds. However, note that we are not taking the {\it exact}
squeezed limit, $k_3\to 0$; rather, we are working in the sub-curvature
approximation, and thus $k_3$ has to satisfy $k_3\gg 1$.
Also, once again, eq.~\eqref{eq:59} is valid at the leading order in the
sub-curvature approximation. See appendix~\ref{sec:tunneling-eval-each} for
sub-leading contributions.

%%%%%%%%%%%%%%%%%%%%%%%%%%%%%%%%%%%
%%%%%%%%%%%%%%%%%%%%%%%%%%%%%%%%%%%
\section{Conclusion}
\label{sec:conclusion}
%%%%%%%%%%%%%%%%%%%%%%%%%%%%%%%%%%%
%%%%%%%%%%%%%%%%%%%%%%%%%%%%%%%%%%%
In this paper, we have computed the bispectrum from a single-field
open inflation model, in which a single scalar field is responsible for both
quantum tunneling (nucleation of a bubble inside of which we live) and
 slow-roll inflation inside a 
bubble after tunneling. We assume that the potential energy inside the
bubble at the onset of slow-roll inflation is approximately equal to the
false vacuum energy 
density of the background de Sitter space.

A comoving observer in de Sitter space charted by open coordinates
detects particles with respect to the Euclidean vacuum state (which is
equivalent to the Bunch-Davies vacuum state in de Sitter space charted
by flat coordinates), in the same
sense that a comoving observer in 
Minkowski space charted by Rindler coordinates detects
particles with respect to the Minkowski vacuum state. Moreover, quantum
tunneling modifies an initial state for 
quantum fluctuations away from the Euclidean vacuum state. Therefore, open
inflation naturally provides a ``non-Bunch-Davies'' initial state for
quantum fluctuations. The occupation number of particles detected inside
the bubble has a thermal spectrum, exponentially falling off at large momenta.

Most of the previous work on the effect of modified
initial states for quantum fluctuations puts mode functions
characterizing a non-Bunch-Davies initial
state by hand at an artificial initial time. Instead of doing this, we
compute the effect of a 
non-Bunch-Davies vacuum on the bispectrum self-consistently within the
framework of open inflation. An ``initial time'' naturally emerges
because slow-roll inflation follows the period of a curvature-dominated
era right after quantum tunneling. The mode functions (hence the initial state)
are uniquely fixed given a model of open inflation.

We find that the bispectrum of the curvature perturbation on a
uniform-density hypersurface, $\zeta$, has a term going as
$B_\zeta(k,k,k_3)\propto e^{-\pi k}k_3^{-4}k^{-2}$ in the squeezed
limit, $k_3\ll k_1\approx k_2\equiv k$. (The units are such that
$k\approx 1$ is a wavenumber corresponding to the curvature radius of
our universe.)
This bispectrum rises more
sharply toward small $k_3$ than that of the standard single-field
inflation model, 
$B_\zeta(k,k,k_3)\propto k_3^{-3}k^{-3}$, by a factor of $k/k_3\gg
1$. This behaviour agrees with phenomenological studies done by the
previous work \cite{Agullo:2010ws,Ganc:2011dy,Chialva:2011hc}. 
However, the amplitude of the bispectrum is exponentially suppressed by
$e^{-\pi k}$ in the sub-curvature region, $k\gg 1$, i.e., for wavelengths
shorter than the present-day curvature radius. Given that we do not see
evidence for spatial curvature within our current horizon, we conclude
that the non-Bunch-Davies effect of open inflation on the observable
bispectrum is exponentially suppressed. This is a consequence of the
occupation number of particles detected in our bubble falling off 
exponentially at large momenta.
We also find that the bispectrum in the exact folded limit, e.g.,
$k_1=k_2+k_3$, remains finite.

The leading-order bispectrum from open inflation in the sub-curvature
approximation is similar to that of the standard single-field model with
a Bunch-Davies initial state.
The bispectrum specific for open inflation arises
only in the sub-leading order in the sub-curvature approximation,
which is suppressed at least by the factor $1/k_i~ (\ll 1)$ compared to the
leading-order bispectrum.

Open inflation provides an attractive framework within which we can discuss
the origin of {\it our} universe without its initial singularity. It is thus
worthwhile to find any observable signatures of open inflation. We have
shown that the bispectrum picks up some signatures of open inflation,
although they are exponentially suppressed in the model we have explored.
On the other hand, it is possible that the false vacuum energy density
is much greater than 
the potential energy inside the bubble. In such a case, the high false
vacuum energy may overcome the exponential suppression factor, $e^{-\pi
k}$. Also, while we have ignored the effect of super-curvature modes
(for which $p^2<0$ where $p$ is the wavenumber of the open harmonics),
they may become important when the false vacuum energy density is high
\cite{Yamauchi:2011qq}. 

\begin{acknowledgments}
KS would like to thank J.~White, D.~Yamauchi, T.~Tanaka and M.~Sasaki
 for useful discussions and valuable comments. KS would also like to
 thank Max-Planck-Institut f\"ur Astrophysik for hospitality, where this
 work was initiated and completed.
This work was supported in part by
 Monbukagaku-sho Grant-in-Aid for the Global COE programs, ``The Next
 Generation of Physics, Spun from Universality and Emergence'' at Kyoto
 University. KS was supported by Grant-in-Aid for JSPS Fellows
 No.~23-3437.
\end{acknowledgments}
\appendix
%%%%%%%%%%%%%%%%%%%%%%%%%%%%%%%%%%%
%%%%%%%%%%%%%%%%%%%%%%%%%%%%%%%%%%%
\section{Expansion of action in a FLRW universe with negative spatial curvature}
\label{sec:expansion}
%%%%%%%%%%%%%%%%%%%%%%%%%%%%%%%%%%%
%%%%%%%%%%%%%%%%%%%%%%%%%%%%%%%%%%%
In this appendix, we derive the reduced action for scalar-type
perturbations by perturbatively expanding the full action, 
which consists of the Einstein-Hilbert action,
the York-Gibbons-Hawking term, and the scalar-field action.
With the metric given in eq.~\eqref{eq:76}, the full action is given by
\begin{align}
 S=\frac{1}{2}\int dtd^3{\bf x} \sqrt{h}&\left\{
N\left[R^{(3)}-2V-h^{ij}\partial_i\phi(t,{\bf x})\partial_j\phi(t,{\bf x})\right]\right.\nnmb
&\left.+\frac{1}{N}\left(E_{ij}E^{ij}-E^2+\left[\dot{\phi}(t,{\bf x})
-N^i\partial_i\phi(t,{\bf x})\right]^2\right)
\right\}\,,\label{eq:2}
\end{align}
where
\begin{align}
 E_{ij}&\equiv \frac{1}{2}\left(\dot{h}_{ij}-\partial_iN_j-\partial_jN_i\right)\,,
\qquad E=h^{ij}E_{ij}\,,
\label{eq:3}
\end{align}
and $R^{(3)}$ is the 3-dimensional Ricci scalar.

In an open FLRW universe, where the curvature of the universe is given
by $K=-1$, the action has a similar form as in the case of a closed FLRW
universe with $K=1$, which has been studied by Clunan and Seery
\cite{Clunan:2009ib}. However, there is one big difference.
In open inflation, where $\dot{\phi}(t)= 0$ and $H(t)\equiv
\dot{a}(t)/a(t)=\infty$ at the initial time, $t=0$,
$\epsilon(\equiv(dV/d\phi)^2/(2V^2))\ll 1$ guarantees
$\dot{\phi}^2/(2H^2)\ll 1$ at all times.
In a closed universe, where $H(t)=0$ at the time of bounce,
$\dot{\phi}^2/(2H^2)\ll 1$ is not satisfied for $\epsilon\ll 1$.
ref.~\cite{Clunan:2009ib} thus imposes a different condition on
$V(\phi)$ in order to satisfy $\dot{\phi}^2/(2H^2)\ll 1$.

%%%%%%%%%%%%%%%%%%%%%%%%%%%%%%%%%%%%%%
\subsection{Second-order action}
\label{sec:second-order-action}
%%%%%%%%%%%%%%%%%%%%%%%%%%%%%%%%%%%%%%
As in section~\ref{sec:mode_open},
taking the uniform curvature (flat) gauge ($\zeta=0$),
we expand $\phi$ around the uniform background, $\phi(t)$, as 
$\phi(t,{\bf x})=\phi(t)+\varphi(t,{\bf x})$, and hereafter denote
$\phi(t)$ as $\phi$. 
We obtain the reduced action by solving the constraint equations after
gauge fixing. For the second- or third-order action, it is enough to
solve the constraint equations up to $O(\varphi)$ \cite{Maldacena:2002vr}. 

By writing $N=1+\delta N$ and $N^i=\gamma^{ij}\partial_j\chi$,
we obtain the Hamiltonian constraint and the momentum constraint
up to $O(\varphi)$, respectively, as
\begin{align}
2 \frac{dV}{d\phi}\varphi +2\delta N (-6H^2+\dot{\phi}^2)
-4H\partial^2\chi-2\dot{\phi}\dot{\varphi}&=0\,,\\
\partial_i\left[
2H\delta N -2K\chi -\dot{\phi}\dot{\varphi}
\right]&=0
\label{eq:10}\,.
\end{align}
From these equations,
we obtain $\delta N$ and $\chi$ up to $O(\varphi)$, respectively, as
 \begin{align}
 \delta N&=\frac{\dot{\phi}}{2H}\varphi+\frac{K}{H}\chi\,,\label{eq:70}\\
 \chi&\approx-\frac{1}{2H}(\partial^2+3K)^{-1}
\left(\dot{\phi}\dot{\varphi}-\ddot{\phi}\varphi\right)\,.
 \label{eq:16}
 \end{align}
Here, we have kept only the leading-order terms in the expansion in
$\dot{\phi}^2/(2H^2)$ in eq.~\eqref{eq:16}.
This expansion is equivalent to 
the expansion in the slow-roll parameter, $\epsilon$. Both $\delta N$
and $\chi$ are $O(\epsilon^{1/2})$ quantities.
We keep the term proportional to $\ddot{\phi}$  eq.~\eqref{eq:16}, as
$\ddot{\phi}/(H\dot{\phi})$ is an $O(1)$ quantity in the
curvature-dominated era. It is however slow-roll suppressed during the
slow-roll inflationary era.

By substituting eqs.~\eqref{eq:70} and \eqref{eq:16} into eq.~\eqref{eq:2}
and keeping quantities up to $O(\varphi^2)$,
we obtain the second-order action as
\begin{align}
 S_2
&=
\int dt d^3{\bf x} a^3\sqrt{\gamma} 
\left[
\left(-\frac{(d^2V/d\phi^2)}{2}-\frac{\dot{\phi}(dV/d\phi)}{2H}\right)
\varphi^2
+\frac{1}{2a^2}\partial_i\varphi\partial^i\varphi
+K\partial_i\chi\partial^i\chi
+\frac{1}{2}\dot{\varphi}^2
\right.\nnmb
&\hspace{3cm}\left.
-\frac{\dot{\phi}^2}{H}\varphi\dot{\varphi}
-\frac{K(dV/d\phi)}{H}\chi\varphi
-\frac{K\dot{\phi}}{H}\chi\dot{\varphi}
+\left(-3+\frac{\dot{\phi}^2}{2H^2}\right)
\left(\frac{\dot{\phi}^2}{4}\varphi^2
+K\dot{\phi}\chi\varphi
+\chi^2
\right)
\right]\nnmb
&\approx\int dtd^3{\bf x}a^3\sqrt{\gamma}
\left[
\frac{1}{2}\dot{\varphi^2}-\frac{1}{2a^2}\partial_i\varphi\partial^i\varphi
\right]\,,
\label{eq:11}
\end{align}
where the sub-leading terms in the slow-roll expansion are neglected in
the last expression. This is eq.~\eqref{eq:47} in section~\ref{sec:review}.

%%%%%%%%%%%%%%%%%%%%%%%%%%%%%%%%%%%
%%%%%%%%%%%%%%%%%%%%%%%%%%%%%%%%%%%
\subsection{Third-order action}
\label{sec:int_by}
%%%%%%%%%%%%%%%%%%%%%%%%%%%%%%%%%%%
%%%%%%%%%%%%%%%%%%%%%%%%%%%%%%%%%%%
We also keep only the leading-order quantities in the slow-roll
expansion in the third-order action. By substituting eqs.~\eqref{eq:70}
and \eqref{eq:16} into eq.~\eqref{eq:2} and keeping quantities up to
$O(\varphi^3)$, we obtain 
\begin{align}
 S_3
&=\int dtd^3{\bf x}\sqrt{\gamma}
\left[
-\left(\frac{a^3\dot{\phi}\varphi}{4H}+
\frac{Ka^3\chi}{2H}\right)
\left(\dot{\varphi}^2+
\frac{1}{a^2}\partial_i\varphi\partial^i\varphi\right)
-a^3\dot{\varphi}\partial_i\chi\partial^i\varphi
\right]\,,
\label{eq:120}
\end{align}
where $S_3$ is an $O(\epsilon^{1/2})$ quantity.

The last term in eq.~\eqref{eq:120} can be integrated by parts to give
 \begin{align}
 -\int dtd^3{\bf x}\sqrt{\gamma}a^3
 \dot{\varphi}\partial_i\chi\partial^i\varphi
 =&\int dtd^3{\bf x}\sqrt{\gamma}
 \left(
 -\frac{3}{2}a^3H\chi\partial_i\varphi\partial^i\varphi
 -\frac{1}{2}a^3\dot{\chi}\partial_i\varphi\partial^i\varphi
 +a^3\chi\dot{\varphi}\partial^2\varphi \right)\,.
 \label{eq:a18}
 \end{align}
The last term in this expression can be integrated by parts
again to give
\begin{align}
\int dtd^3{\bf x}\sqrt{\gamma}
a^3\chi\dot{\varphi}\partial^2\varphi
&=\int dtd^3{\bf x}\left\{\left[\sqrt{\gamma}
\left(\frac{a^5H}{2}\chi\dot{\varphi}^2-
\frac{a^5}{2}\dot{\chi}\dot{\varphi}^2\right)
\right]
+a^2\chi \dot{\varphi}\frac{\delta {{\cal L}}}{\delta \varphi}\right\}\,,
\label{eq:64}
\end{align}
where the Lagrangian density, ${\cal L}$, is defined by the
second-order action as $S_2=\int
d^4x~{\cal L}$.
Substituting this back into eq.~\eqref{eq:a18} we get
\begin{align}
 -\int dtd^3{\bf x}\sqrt{\gamma}a^3
 \dot{\varphi}\partial_i\chi\partial^i\varphi
 =&\int dtd^3{\bf x}
\left\{\left[\sqrt{\gamma}
 \left(
\frac{a^5H}{2}\chi
\left(\dot{\varphi}^2-3\partial_i\varphi\partial^i\varphi\right)
 -\frac{a^5}{2}\dot{\chi}
\left(\dot{\varphi}^2+\frac{1}{a^2}\partial_i\varphi\partial^i\varphi\right)
\right)\right]\right.\nnmb
&\qquad\qquad\left.+a^2\chi
\dot{\varphi}\frac{\delta {{\cal L}}}{\delta \varphi}\right\}\,.
\label{eq:9}
\end{align}
The time derivative of $\chi$ can be written as
 \begin{align}
 \dot{\chi}
 &=-\frac{K}{a^2H}\chi
 -3H\chi-\frac{\dot{\phi}}{2Ha^2}\varphi
+\frac{3K\dot{\phi}}{2Ha^2}(\partial^2+3K)^{-1}\varphi
 +\frac{\dot{\phi}}{2H}(\partial^2+3K)^{-1}
 \frac{\delta L}{\delta \varphi}\,,
 \label{eq:119}
 \end{align}
where $\delta L/\delta\varphi$ is given in eq.~\eqref{eq:60}. ($L$
is related to ${\cal L}$ by ${\cal L}=a^3\sqrt{\gamma}L$.)
We have ignored the term proportional to $\dddot{\phi}$ using $|\dddot{\phi}/(H^2\dot{\phi})|\ll 1$, which can be derived from eq.~\eqref{eq:36}.
By substituting eq.~\eqref{eq:119} into eq.~\eqref{eq:9}, we obtain
 \begin{align}
 &-\int dtd^3{\bf x}\sqrt{\gamma}a^3
 \dot{\varphi}\partial_i\chi\partial^i\varphi\nnmb
&=\int dtd^3{\bf x}\left\{\left[\sqrt{\gamma}\left(
2a^5H\chi\dot{\varphi}^2
+\left(
\frac{Ka^3\chi}{2H}
+\frac{a^3\dot{\phi}}{4H}\varphi
-\frac{3Ka^3\dot{\phi}}{4H}(\partial^2+3K)^{-1}\varphi
\right)
\left(\dot{\varphi}^2+\frac{1}{a^2}\partial_i\varphi\partial^i\varphi \right)
\right)\right]\right.\nnmb
&\hspace{5.5cm}
\left.
+\left(
-\frac{a^2\dot{\phi}}{4H}
\left(\dot{\varphi}^2+\frac{1}{a^2}\partial_i\varphi\partial^i\varphi \right)
(\partial^2+3K)^{-1}
+a^2\chi\dot{\varphi}\right)
\frac{\delta {\cal L}}{\delta \varphi}
\right\}\,.
 \label{eq:210}
\end{align}

Finally, substituting eq.~\eqref{eq:210} into eq.~\eqref{eq:120},
we obtain the third-order action
\begin{align}
 S_3&=
\int dtd^3{\bf x}\sqrt{\gamma}\left[
-a^5(\partial^2+3K)^{-1}\left(\dot{\phi}\dot{\varphi}-\ddot{\phi}\varphi\right)
\dot{\varphi}^2
-\frac{3Ka^3\dot{\phi}}{4H}(\partial^2+3K)^{-1}\varphi
\left(\dot{\varphi}^2
+\frac{1}{a^2}\partial_i\varphi\partial^i\varphi\right)\right]\nnmb
&+\int dtd^3{\bf x}
\left[-\frac{a^2\dot{\phi}}{4H}
\left(\dot{\varphi}^2+\frac{1}{a^2}\partial_i\varphi\partial^i\varphi \right)
(\partial^2+3K)^{-1}
-\frac{a^2}{2H}(\partial^2+3K)^{-1}\dot{\varphi}
\left(\dot{\phi}\dot{\varphi}
-\ddot{\phi}\varphi\right)\right]
\frac{\delta {\cal L}}{\delta \varphi}\,.
\end{align}
This is eq.~\eqref{eq:29} in section~\ref{sec:ng-gen} after moving
$(\partial^2+3K)^{-1}$ in the first term in the second line to the left
by integration by parts, and replacing $K$ with $K=-1$.

%%%%%%%%%%%%%%%%%%%%%%%%%%%%%%%%%%%
%%%%%%%%%%%%%%%%%%%%%%%%%%%%%%%%%%%
\section{Analytical structure of the universe in open inflation}
\label{sec:tunneling-region}
%%%%%%%%%%%%%%%%%%%%%%%%%%%%%%%%%%%
%%%%%%%%%%%%%%%%%%%%%%%%%%%%%%%%%%%

%%%%%%%%%%%%%%%%%%%%%%%%%%%%%%%%%%%
%%%%%%%%%%%%%%%%%%%%%%%%%%%%%%%%%%%
\subsection{Coordinate systems and Wightman functions}
%%%%%%%%%%%%%%%%%%%%%%%%%%%%%%%%%%%
%%%%%%%%%%%%%%%%%%%%%%%%%%%%%%%%%%%
The coordinate system defined by eqs.~\eqref{eq:127-E} to
\eqref{eq:127-C} is suitable for studying one of the mode functions
forming a complete set, $u_{\bf k}(x)$, which multiplies the annihilation
operator, $\hat{a}_{\bf k}$, in eq.~\eqref{eq:57}. Let us call this
coordinate system ``$K_2$.'' The $K_2$ is suitable for $u_{\bf k}(x)$
because the $R$-, $L$- and $C$-regions are connected via $E_2$, and
$u_{\bf k}(x)$ is analytical in and across all the regions. 

On the other hand, $K_2$ is {\it not} suitable for studying another
mode function, $v_{\bf k}(x)$, which multiplies the creation operator,
$\hat{a}_{\bf k}^\dagger$, in eq.~\eqref{eq:57}. We thus introduce a new
set of coordinate systems, which we shall call ``$K_1$.'' The $K_1$ is
defined by
\begin{align}
& \begin{cases}
  \tau&\left(-\pi/(2H_I)\leq \tau\leq\pi/(2H_I)\right)\\
 \chi&\left(0\leq \chi\leq\pi\right)\hspace{0.2cm},
 \end{cases}\qquad
 \begin{cases}
  t_R= -i(\tau-\pi/(2H_I))&\left(0\leq t_R<\infty\right)\\
 r_R=i(\chi-\pi)&\left(0\leq r_R<\infty\right),\hspace{0.2cm} 
 \end{cases}\nnmb
&\begin{cases}
 t_L=-i\left(-\tau-\pi/(2H_I)\right)&\left(0\leq t_L<\infty\right)\\
 r_L=i(\chi-\pi)&\left(0\leq r_L<\infty\right),\hspace{0.2cm} 
\end{cases}
\begin{cases}
 t_C=\tau&\left(-\pi/(2H_I)\leq t_C\leq \pi/(2H_I)\right)\\
 r_C=i\left(\chi-\pi/2\right)&\left(0\leq r_C<\infty\right)\,,
\end{cases}
\label{eq:128}
\end{align}
with the coordinates of a 2-sphere $\Omega^i$ commonly used for all regions.
For simplicity, we assume that the spacetime is approximately de Sitter
with the metric given by eq.~\eqref{eq:103}. In the $K_1$ system, the $R$-,
$L$- and $C$-regions are connected via $E_1$, and $v_{\bf k}(x)$ is
analytical in and across all the regions.
The difference between $K_2$ and $K_1$ is not a mere coordinate transformation,
but they define the relations between the $E$-, $R$-, $L$-, and $C$-regions in a different way. 

The two Wightman functions coincide for space-like separated $x$ and
$x'$, i.e., $G^{+}(x,x')=G^{-}(x,x')~(\equiv G(x,x'))$, where
$G^+(x,x')=\sum_{\bf k}u_{\bf k}(x)v_{\bf k}(x')$ and
$G^-(x,x')=\sum_{\bf k}u_{\bf k}(x')v_{\bf k}(x)$; however, they are different
for time-like separated $x$ and $x'$.
The Wightman functions diverge
when $x$ is on the past- or future-light cones of  $x'$.
This divergence is the so-called UV divergence of the 
Wightman functions.

In performing calculations it is necessary to choose
one of the two sets of coordinate systems, $K_2$ or $K_1$,
and to choose one of the two Wightman functions, $G^{+}(x,x')$ or $G^{-}(x,x')$.
We shall discuss the origin of these possible options,
in terms of the analytical structure of de Sitter space charted by
these coordinates.
We shall then explain as to why $u_{\bf k}(x)$ is analytical in $K_2$, 
whilst $v_{\bf k}(x)$ is not; and why $v_{\bf k}(x)$ is
analytical in $K_1$, whilst $u_{\bf k}(x)$ is not \cite{Yamamoto:1996qq}.

To discuss the analytical structure of de Sitter space charted by
these coordinates, let us chart the 5-dimensional Minkowski spacetime,
$ds_5^2=\eta_{ab}dx^adx^b$ with $\eta={\rm diag}(-1,1,1,1,1)$,
by the Cartesian coordinates 
\begin{align}
 x=(x^0,x^1,x^2,x^3,x^4)\,.
\label{eq:33}
\end{align}
The 4-dimensional de Sitter spacetime is embedded as a hyperboloid defined by
\begin{align}
 -(x^0)^2+\sum_{i=1}^4(x^i)^2=\frac{1}{H_I^2}\,.
\end{align}
The Euclidean region is given by imposing $x^0$ to be pure imaginary,
and it is related to the coordinates given in
eqs.~\eqref{eq:127-E} to \eqref{eq:127-C} or
eq.~\eqref{eq:128} as \cite{Sasaki:1994yt}
\begin{align}
 ix^0=\cos\tau\cos\chi\,,\quad x^1=\sin\tau\,,\quad
\begin{pmatrix}
 x^2\\x^3\\x^4
\end{pmatrix}
=\cos\tau\sin\chi 
\,{\bf \Omega}\,,
\end{align}
where ${\bf \Omega}=(\Omega^1,\Omega^2,\Omega^3)$ is a 3-dimensional
unit vector. 
With the Cartesian coordinates,
we can discuss the analytical structure of de Sitter
spacetime without worrying about coordinate singularities.

In QFT on an open inflation background,
we encounter two types of singularities.
One is a coordinate singularity and
the other is a UV divergence of the Wightman functions.
The first issue is the coordinate singularity.  Let us consider the
coordinates $(t_C,r_C,\Omega)$ defined in the $C$-region
(i.e., $(-\pi/(2H_I)\leq t_C\leq \pi/(2H_I)$ and $0\leq r_C<\infty$) with
the metric given by the fourth line of
eq.~\eqref{eq:103}. The coordinates in
the $C$-region are analytical\footnote{There is no difference between $K_1$ and
$K_2$ in the $C$-region.} (i.e., the metric is analytical and
non-singular), which means that even if the coordinates are given only
in a finite part of the $C$-region, we can analytically extend them to the
whole $C$-region. However, coordinate singularities appear on the
boundaries between the $C$- and $R$-regions and the $C$- and $L$-regions, and
the analytical extension of the coordinate system beyond these
boundaries seems not possible, even though the spacetime is extended.

\begin{figure}
 \begin{minipage}{0.49 \hsize}
 \includegraphics[width=8cm]{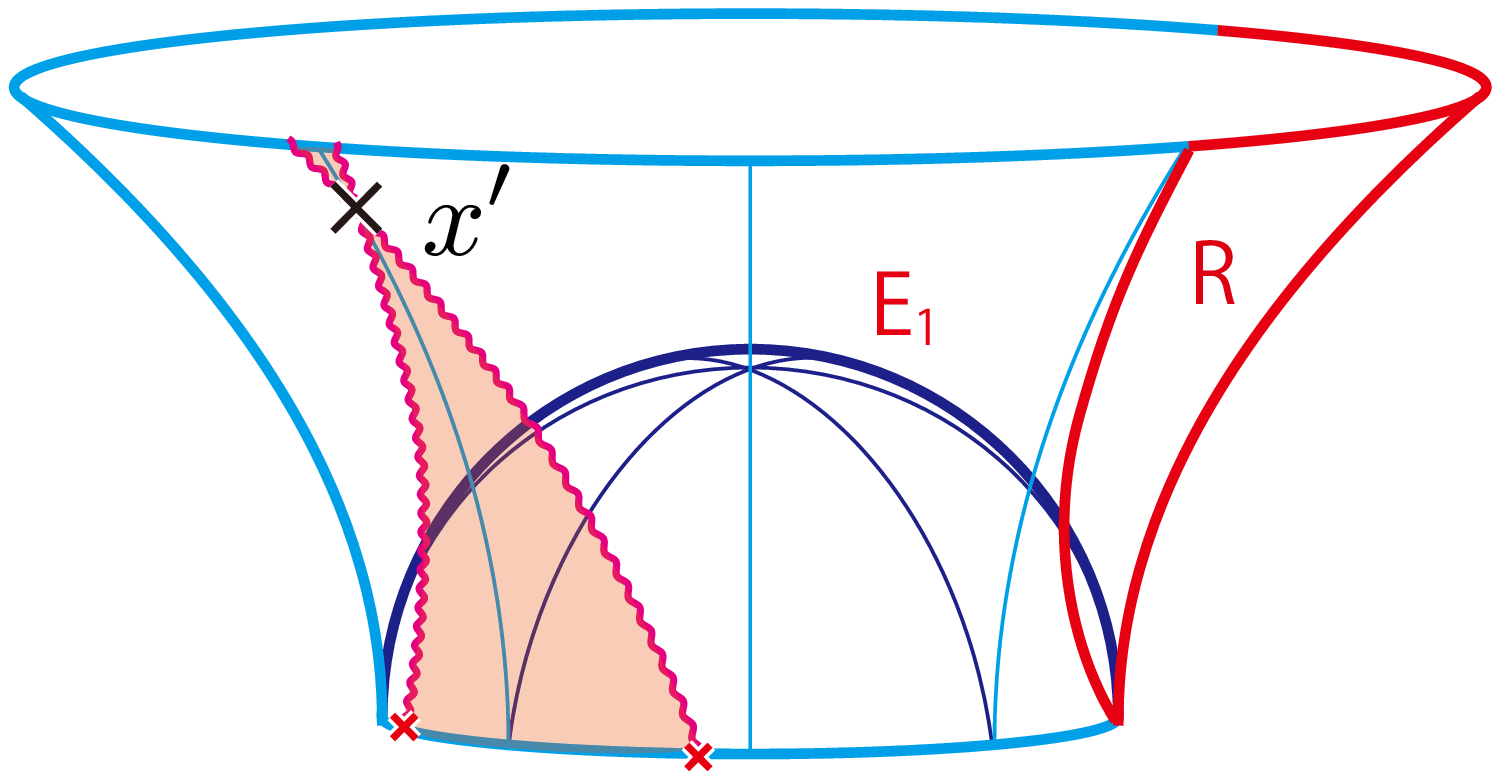}
 \end{minipage}
 \begin{minipage}{0.49 \hsize}
 \includegraphics[width=8cm]{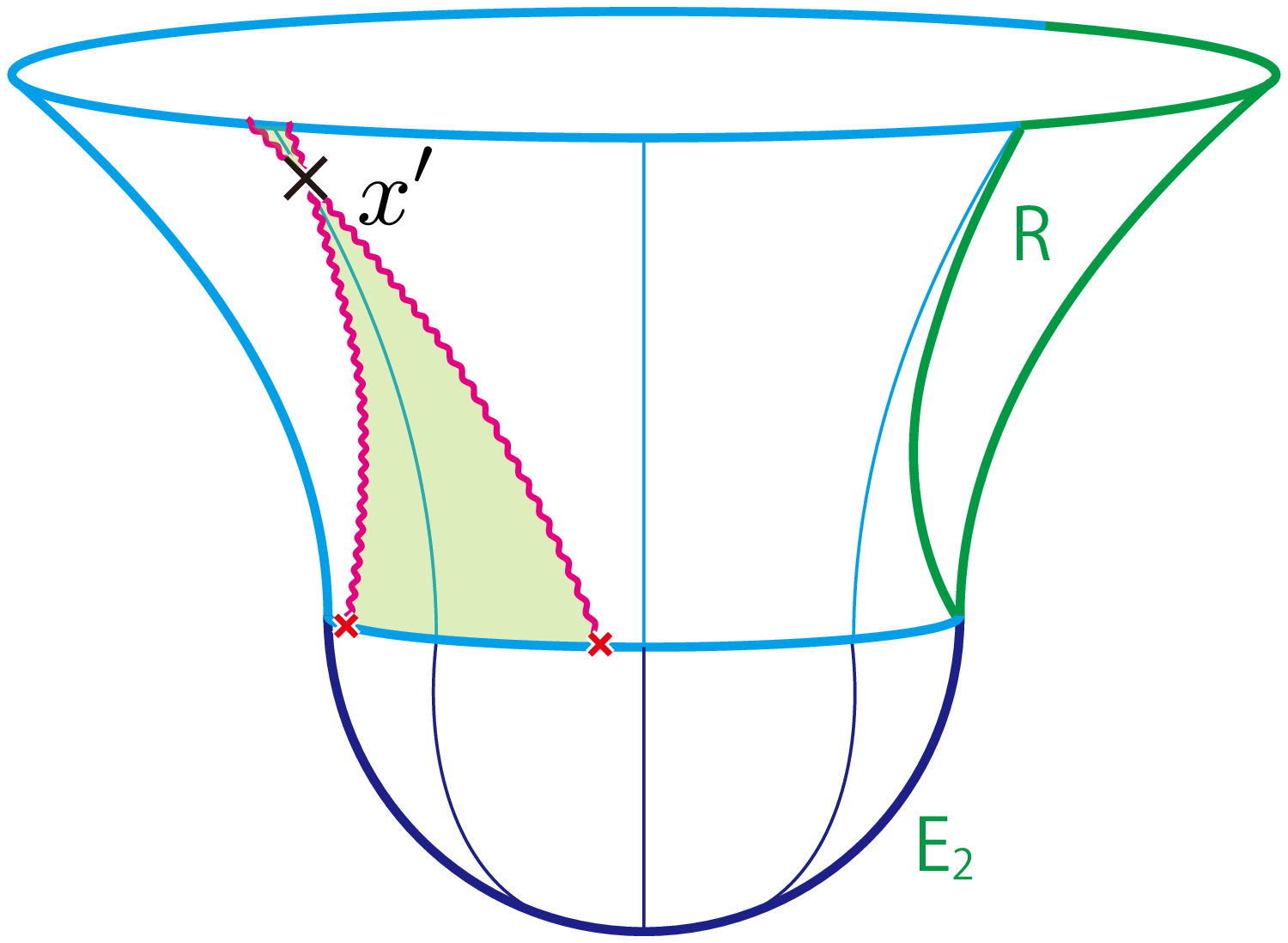}
 \end{minipage}
\caption{Analytical continuation. (Left) The $R$-region is connected to
the $C$-region via $E_1$ by $x^+$, where $\im x^0>0$.
(Right) The $R$-region is connected to
the $C$-region via $E_2$ by $x^-$, where $\im x^0<0$.
The Wightman function, $G(x,x')$, is uniquely determined in the
 space-like region of $x'$, but it diverges on the light-cone of $x'$
 shown by the wavy lines. Analytically extending $G(x,x')$ into the
 past-light cone of $x'$ using $x^+$ and $x^-$, we obtain $G^{-}(x,x')$
 in the shaded  area in the left panel and $G^{+}(x,x')$ in the shaded
 area in the right panel, respectively.} 
\label{fig:5d_ann} 
\end{figure}

The second issue is the UV divergence of the Wightman functions.
Let us fix $x'$ and 
regard the Wightman function, $G(x,x')$, as a function of $x$.
As $G(x,x')$ is analytical in the space-like region of $x'$,
we can analytically extend it to the whole space-like region of $x'$,
even if $G(x,x')$ is given only in a finite part of the space-like
region of $x'$.
However, $G(x,x')$ diverges on the past- or future-light cones of $x'$,
and again, the analytical extension of $G(x,x')$ beyond the past- or future-light cones of $x'$ seems not possible.

Those apparent issues regarding the analytical extension
of the coordinates and the Wightman functions can be overcome by taking
a path through the complex plane. 
By adding an infinitely small complex number, $\pm i\epsilon(\epsilon>0)$,
to $x^0$, we obtain 
\begin{align}
x^{\pm}=(x^0\pm i\epsilon,x^1,x^2,x^3,x^4)\,.
\end{align}
Since $x^{\pm}$ no longer passes right through the singularities,
the coordinates and the Wightman functions can be analytically extended
to all the regions charted by $K_1$ or $K_2$.

Let us take $x^{+}$ first, and see how the analytical extension 
is performed.
The coordinate system defined in the $C$-region
can be analytically extended with $x^{+}$
beyond the coordinate singularities on the boundaries
between the $C$- and $R$-regions and the $C$- and $L$-regions.
As $\im x^0 >0$ in $E_1$, the resulting set of coordinate systems
corresponds to $K_1$, which connects the $R$-, $L$-, and $C$-regions via
$E_1$ (see figure~\ref{fig:5d_ann}). 
The Wightman function $G(x,x')$, considered as a function of $x$ for a
given $x'$ ($x$ is in the space-like region of $x'$), can be
analytically extended by shifting $x^+$ beyond the past-light cone of
$x'$, where $G(x,x')$ diverges (see section~5.4 of ref.~\cite{birrell-davies}).
As a result, $G^{-}(x,x')$ is obtained inside the past-light cone of
$x'$ as shown in
figure~\ref{fig:5d_ann}. 
Since the analyticity of $v_{\bf k}(x)$ and $G^{-}(x,x')$ with respect to $x$
are equivalent, 
the analyticity of $v_{\bf k}(x)$ with respect to $x^{+}$ is also guaranteed,
which leads to the analyticity of $v_{\bf k}(x)$ with respect to $K_1$ and the 
regularity of $v_{\bf k}(x)$ on $E_1$.

Similarly, the analytical extension using $x^{-}$ gives $K_2$,
which connects the $R$-, $L$-, and $C$-regions via $E_2$. 
As a result, $G^{+}(x,x')$ is obtained inside the past-light cone of
$x'$. Then $u_{\bf k}(x)$ is
analytical with respect to $K_2$ and is regular on $E_2$.

%%%%%%%%%%%%%%%%%%%%%%%%%%%%%%%%%%%
%%%%%%%%%%%%%%%%%%%%%%%%%%%%%%%%%%%
\subsection{Deformation of integration path using analyticity}
%%%%%%%%%%%%%%%%%%%%%%%%%%%%%%%%%%%
%%%%%%%%%%%%%%%%%%%%%%%%%%%%%%%%%%%
\begin{figure}
\begin{center}
 \begin{minipage}{0.45 \hsize}
 \includegraphics[width=7cm]{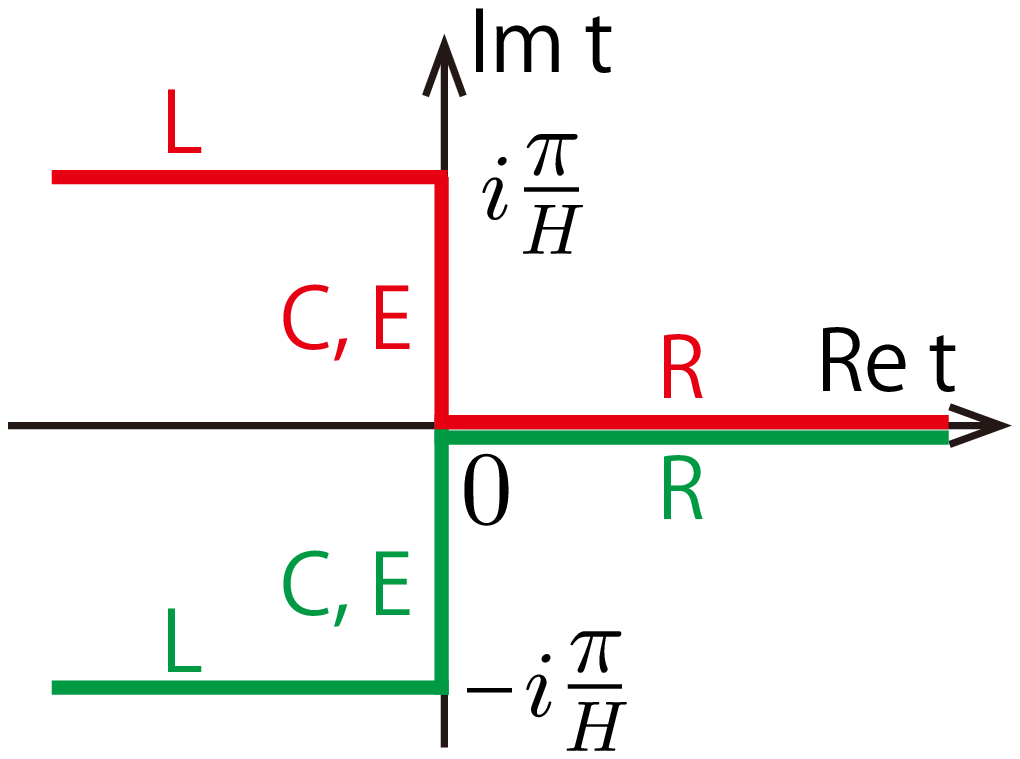}
\caption{Correspondence between the domains of $t$ and
the $E$-, $R$-, $L$-, and $C$-regions.
The upper (red) lines show the integration path $V_1$ charted by $K_1$},
while the bottom (green) lines show $V_2$ charted by $K_2$.
\label{fig:tau}
 \end{minipage}\qquad
 \begin{minipage}{0.45 \hsize}
 \includegraphics[width=7cm]{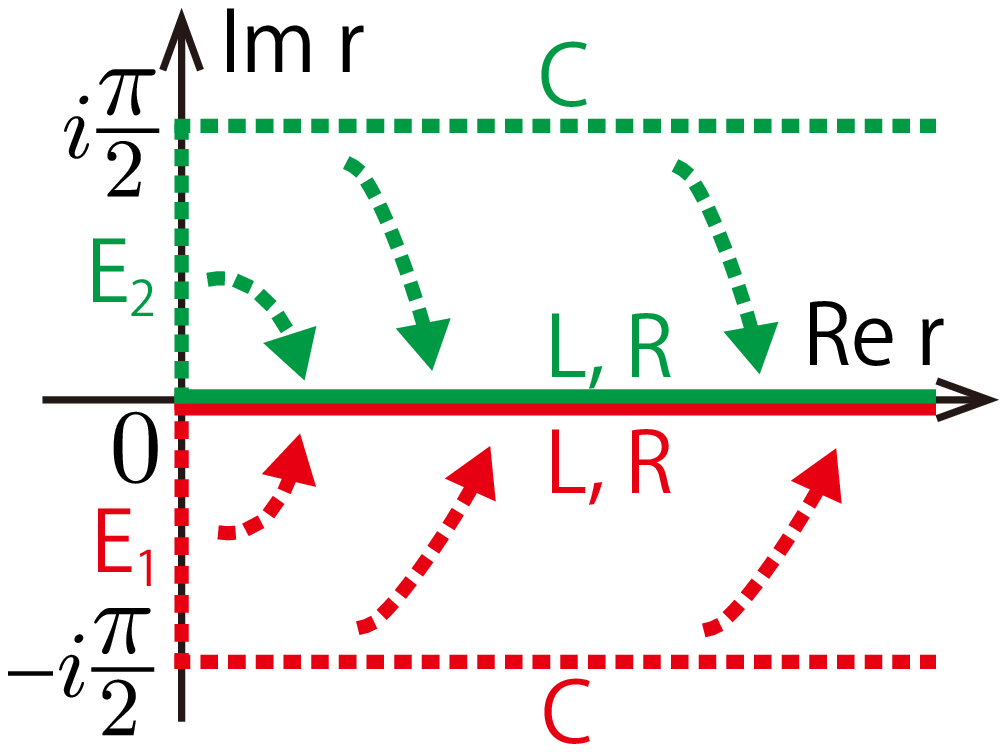}
\caption{Same as figure~\ref{fig:tau}, but for $r$. The bottom (red)
  dashed lines show the original integration path $V_1$ 
  charted by $K_1$,  while the upper (green) dashed lines show $V_2$
  charted by $K_2$. The integral can be evaluated by deforming the paths to lie
  along the real axis of $r$, as shown by the solid lines.
}
\label{fig:chi}
 \end{minipage}
\end{center}
\end{figure}

Thanks to the analyticity of $u_{\bf k}(x)$ and $v_{\bf k}(x)$
with respect to the set of coordinate systems $K_2$ and $K_1$, respectively,
we can deform the path of integration given in eq.~\eqref{eq:24}.
For simplicity, we shall consider the tree-level calculation 
and write the integral schematically as 
$\int_{(C_1+C_2)\times \Sigma_\lambda} d\lambda d^3{\bf x} \sqrt{-g}f(x)$.

Let us consider the integral over $V_1=C_1\times \Sigma_\lambda$
in the coordinate system $K_1$.
We can use  the coordinates in the $R$-region, $t$ and $r$, 
in the whole $V_1$ by allowing $t$ and $r$ to be complex variables.
Using the relation between coordinates given by eq.~\eqref{eq:128},
$V_1$ is given by the sum of
\begin{align}
 E_1&=\left\{(t,r,\Omega)|t\in(0,i\pi/H_I),r\in(0,-i\pi/2),\Omega\in S_2\right\}\,,\nnmb
 C&=\left\{(t,r,\Omega)|t\in(0,i\pi/H_I),r\in(-i\pi/2,-i\pi/2+\infty),\Omega\in S_2\right\}\,,\nnmb
 R&=\left\{(t,r,\Omega)|t\in(0,\infty),r\in(0,\infty),\Omega\in S_2\right\}\,,\nnmb
 L&=\left\{(t,r,\Omega)|t\in(i\pi/H_I,-\infty),r\in(0,\infty),\Omega\in S_2\right\}\,,
\label{eq:34}
\end{align}
where, for simplicity, we additionally assume that $\Sigma_0$ is in the 
far future of the nucleation surface and that
the Lorentzian region of $V_1$ can be approximated as the whole
de Sitter spacetime after the nucleation time.
We can then simplify the integration domain by deforming the integration path as
\begin{align}
 \int_{C_1\times \Sigma_\lambda} d\lambda d^3{\bf x} \sqrt{-g}f(x)
&=\int_{i\pi/H_I-\infty}^\infty dt \int_0^\infty dr  \int d\Omega \sqrt{-g}f(x)\,.
\label{eq:4}
\end{align}
The upper (red) lines in figure~\ref{fig:tau} show the integration
path along $t$, while the bottom (red) lines in figure~\ref{fig:chi}
show how we deform the integration path to lie along the real axis of $r$.
The deformation shown in figure~\ref{fig:chi} is possible because 
the integrand $\sqrt{-g}f(x)$ (which is a function of the background
quantities and $v_{\bf k}(x)$, both of which are analytical) is
analytical with respect to $t$ and $r$ 
given by eq.~\eqref{eq:34}, and $\sqrt{-g}f(x)$ falls off rapidly as
$\re r\to\infty$ for sub-curvature modes, i.e., $p^2>0$ for which the
open harmonics falls off as $f_{pl}(r)\propto e^{-r}$ (see
eq.~\eqref{eq:30}). However, this deformation would not be valid for
super-curvature modes, which we ignore in this paper.

Similarly, with the set of coordinate systems $K_2$ given by
eqs.~\eqref{eq:127-E} to \eqref{eq:127-C}, the domain of integration
$V_2=C_2\times\Sigma_\lambda$ is given by the sum of 
\begin{align}
 E_2&=\left\{(t,r,\Omega)|t\in(0,-i\pi/H_I),r\in(0,i\pi/2),\Omega\in S_2\right\}\,,\nnmb
 C&=\left\{(t,r,\Omega)|t\in(0,-i\pi/H_I),r\in(i\pi/2,i\pi/2+\infty),\Omega\in S_2\right\}\,,\nnmb
 R&=\left\{(t,r,\Omega)|t\in(0,\infty),r\in(0,\infty),\Omega\in S_2\right\}\,,\nnmb
 L&=\left\{(t,r,\Omega)|t\in(-i\pi/H_I,-\infty),r\in(0,\infty),\Omega\in S_2\right\}\,,
\label{eq:71}
\end{align}
and we can simplify the integration domain by deforming the integration path as
\begin{align}
 \int_{C_2\times \Sigma_\lambda} d\lambda d^3{\bf x} \sqrt{-g}f(x)
&=-\int_{-i\pi/H_I-\infty}^\infty dt \int_0^\infty dr
 \int d\Omega \sqrt{-g}f(x)\,,
\label{eq:48}
\end{align}
as is shown in the bottom (green) lines in figure~\ref{fig:tau} and the
upper (green) lines in figure~\ref{fig:chi}.

In the single-field open inflation model that we study in this paper,
the mode functions contributing to the bispectrum in the $R$-region
vanish outside the bubble. 
Thus, eqs.~\eqref{eq:4} and \eqref{eq:48} further simplify to
\begin{align}
 \int_{C_1\times \Sigma_\lambda} d\lambda d^3{\bf x} \sqrt{-g}f(x)
&=\int_{iR_W}^\infty dt \int_0^\infty dr \int d\Omega 
\sqrt{-g}f(x)\,,\label{eq:75}\\
\int_{C_2\times \Sigma_\lambda} d\lambda d^3{\bf x} \sqrt{-g}f(x)
&=-\int_{-iR_W}^\infty dt \int_0^\infty dr \int d\Omega 
\sqrt{-g}f(x)\,,
\label{eq:19}
\end{align}
respectively, where $R_W$ is the bubble radius measured in the
coordinates of the $E$-regions.

%%%%%%%%%%%%%%%%%%%%%%%%%%%%%%%%%%%
%%%%%%%%%%%%%%%%%%%%%%%%%%%%%%%%%%%
\section{Order-of-magnitude estimate for sub-leading contributions}
\label{sec:tunneling-eval-each}
%%%%%%%%%%%%%%%%%%%%%%%%%%%%%%%%%%%
%%%%%%%%%%%%%%%%%%%%%%%%%%%%%%%%%%%
In section~\ref{sec:sq_ng}, we have computed the leading-order
contribution to the bispectrum, $\mathcal{L}^{(1)}_{int}$, in the
$R$-region, but have ignored the other contributions.
In this appendix, we show that the contributions from the other terms in
$\mathcal{L}_{int}$ are sub-dominant in the sub-curvature approximation.
We also show that the contributions from outside the $R$-region are
sub-dominant in the sub-curvature approximation, which implies that the
effects of possibly large self-interaction near the bubble wall are also
sub-dominant. 
For simplicity, we shall show the result only in the squeezed
configuration, $k_3\ll k_1\approx k_2~(\equiv k)$.

In the following calculations, we change the integration variable
from $t$ to the conformal time, $\eta$, in the integration given in
eq.~\eqref{eq:19}.  
Accordingly, the integration domain changes from $t:-iR_W\to 0\to \infty$
to $\eta:\eta_W\to -\infty \to 0$, where $\eta_W$ is given by
\begin{align}
\eta_W\equiv
-\frac{1}{2}\log\left(\frac{\cosh H_It+1}{\cosh H_It-1}\right)\Bigg|_{t=-iR_W}
\qquad
\left(\quad\Leftrightarrow \quad
e^{\eta_W}=\left(\frac{H_IR_W}{\sqrt{2}}\right)e^{-i\pi/2}\right)\,.
\label{eq:78}
\end{align}
The approximate forms of $a$, $H$, $\dot{\phi}$, $\ddot{\phi}$, and $u_k$
given in eqs.~\eqref{eq:18}, \eqref{eq:36}, and \eqref{eq:31} originally
obtained with the condition $-\infty< \eta\lesssim -1$ can also be used
when $\eta\in (\eta_W, -\infty)$.
This is because the approximate forms for $-\infty< \eta\lesssim -1$
(or $0<t\lesssim H_I^{-1}$) are obtained using $|tH_I|\ll 1$, which is
also satisfied for $\eta\in (\eta_W, -\infty)$ (or $t\in (-iR_W,0)$), as
the radius of the nucleated bubble is smaller than the Hubble scale
(i.e., $H_IR_W\lesssim 1$). 
Throughout this appendix, we shall ignore the effects of the non Bunch-Davies
vacuum state, as we show in section~\ref{sec:sq_ng} that 
these effects are sub-dominant in
the sub-curvature approximation.

\subsection{Contributions from outside the $R$-region}
We calculate the contribution to the leading-order bispectrum,
$B^{(1)}(k_1,k_2,k_3)$, from $\mathcal{L}^{(1)}_{int}$ outside the
$R$-region, $\eta\in(\eta_W,-\infty)$. 
Using the same methods as were used to obtain eq.~\eqref{eq:38}, 
$B^{(1)}(k_1,k_2,k_3)$ is given by
\begin{align}
 B^{(1)}(k_1,k_2,k_3)
 =
 2\re\left[iv_{k_1}(0)v_{k_2}(0)v_{k_3}(0)
 \left(\int_{\eta_W}^{-\infty} d\eta  
 \frac{a^6\dot{\phi}}{k_1^2+4}
\dot{u}_{k_1}(\eta)\dot{u}_{k_2}(\eta)\dot{u}_{k_3}(\eta)\right)
+{\rm (perms.)} 
\right]\,,
\label{eq:77}
\end{align}
where only the integration domain is different from eq.~\eqref{eq:38}.
Substituting eqs.~\eqref{eq:36}, \eqref{eq:31}, \eqref{eq:44}, and
\eqref{eq:78} into eq.~\eqref{eq:77}, we obtain
\begin{align}
B^{(1)}(k_1,k_2,k_3)
&=
 2\re\left[i\left(\frac{H_I^3}{k_1^2}\right)
\frac{1}{8k_1k_2k_3}
\int_{\eta_W}^{-\infty}d\eta \dot{\phi}
e^{-i(k_1+k_2+k_3)\eta}
+{\rm (perms.)} \right]\nnmb
&\approx
\frac{\sqrt{\epsilon}H_I^4e^{-\pi(2k+k_3)/2}}{k_3^3k^3}H_IR_W\,,
\label{eq:22}
\end{align}
where we do not show an $O(1)$ pre-factor, which does not affect the
order-of-magnitude estimate. 

The exponential suppression factor, $e^{-\pi(2k+k_3)/2}$, appears
because the mode functions of sub-curvature modes are
exponentially suppressed outside the $R$-region. 
(One may check this by substituting $\eta\in(\eta_W,-\infty)$ into
$u_k(\eta)$ given in eq.~\eqref{eq:37}.) This factor makes
eq.~\eqref{eq:22} much smaller than $B(k_1,k_2,k_3)$ in
eq.~\eqref{eq:40} in the sub-curvature approximation.
This suppression is universal for any integrations outside the $R$-region.
We thus conclude that the integration of all terms in $\mathcal{L}_{int}$
outside the $R$-region is negligible in the sub-curvature
approximation. This also implies that the contribution from the possibly
large self-interaction near the bubble wall (which is in the
$C$- and $E$-regions) is negligible in the sub-curvature approximation. 

\subsection{Contributions from the sub-leading terms in  Lagrangian}
Next, we calculate the contributions from the other terms of
$\mathcal{L}_{int}$ given in eq.~\eqref{eq:39}. In addition to the
leading-order term, 
$\mathcal{L}_{int}^{(1)}$, there are three terms:   
\begin{align}
\mathcal{L}_{int}^{(2)}=&
\sqrt{\gamma}
a^5\ddot{\phi}(\partial^2-3)^{-1}\varphi_{c}\dot{\varphi_{c}}^2\,,\nnmb
\mathcal{L}_{int}^{(3)}=&
 \sqrt{\gamma}\frac{3a^3\dot{\phi}}{4H}(\partial^2-3)^{-1}\varphi_{c}
 \dot{\varphi_{c}}^2\,,\nnmb
\mathcal{L}_{int}^{(4)}= &-\sqrt{\gamma}
\frac{3a^3\dot{\phi}}{4H}(\partial^2-3)^{-1}\varphi_{c}
 \frac{1}{a^2}\partial_i\varphi_{c}\partial^i\varphi_{c}\,.
\label{eq:56}
\end{align}
As the integral outside the $R$-region is negligible
in the sub-curvature approximation, we shall consider the contributions
from the integral inside the $R$-region. 

Let us start with $\mathcal{L}_{int}^{(2)}$. The relative size of
$\mathcal{L}_{int}^{(2)}$ compared to $\mathcal{L}_{int}^{(1)}$
is $\mathcal{L}_{int}^{(2)}/\mathcal{L}_{int}^{(1)}\approx
\ddot{\phi}\varphi_c/\dot{\phi}\dot{\varphi_c}$. 
According to eqs.~\eqref{eq:36}, \eqref{eq:31}, and \eqref{eq:44}, this
ratio is $\ddot{\phi}\varphi_c/\dot{\phi}\dot{\varphi_c}\approx aH/k$
for $-\infty<\eta\lesssim -1$, and
$\ddot{\phi}\varphi_c/\dot{\phi}\dot{\varphi_c}\approx O(\epsilon)$ for
$-1\lesssim \eta <0$.
The ratio in $-1\lesssim \eta <0$ is slow-roll suppressed.
The ratio in  $-\infty<\eta\lesssim -1$ is small for sub-horizon modes with
$k/(aH)\gg 1$. The horizon size and the curvature radius are
comparable, i.e., $aH\approx 1$, during the curvature-dominated era,
$-\infty<\eta\lesssim -1$. Therefore, the ratio in
$-\infty<\eta\lesssim -1$ is suppressed in the sub-curvature
approximation, $k\gg 1$.

Next, we calculate the contribution from $\mathcal{L}_{int}^{(3)}$. The
bispectrum is 
\begin{align}
B^{(3)}(k_1,k_2,k_3)=
  2\re&\left[iu_{k_1}(0)u_{k_2}(0)u_{k_3}(0)
 \left(\int_{-\infty}^0d\eta
 \frac{a^4\dot{\phi}}{4H(\eta)(k_1^2+4)}
u_{k_1}(\eta)\dot{u}_{k_2}(\eta)\dot{u}_{k_3}(\eta)\right)\right.\nnmb
&+{\rm (perms.)} \Big]\,.
\label{eq:6}
\end{align}
Using eqs.~\eqref{eq:18}, \eqref{eq:36}, \eqref{eq:31}, and
\eqref{eq:44}, we estimate the contribution from  $\eta\in(-1,0)$ as
\begin{align}
{\sqrt{\epsilon}}H^4\,
 \re\left[i
 \frac{1}{k_1^5k_2k_3}
\int_{-1}^{0}d\eta~
(1+ik_1\eta)e^{-i(k_1+k_2+k_3)\eta}
\right] +{\rm (perms.)} 
\approx \frac{\sqrt{\epsilon}H^4}{k_3^4k^3}\,,
\label{eq:45}
\end{align}
where the integral is estimated as $O(k_1/k_T)$
with $k_T\equiv k_1+k_2+k_3$.
Using eqs.~\eqref{eq:18}, \eqref{eq:36}, \eqref{eq:31}, and
\eqref{eq:44}, we estimate the contribution from $\eta\in(-\infty,-1)$ as
\begin{align}
\frac{\sqrt{\epsilon}H^4}{k_1^4k_2k_3}
 \re\left[
\int_{-\infty}^{-1}d\eta~
e^{\eta}
e^{-i{k_1}\eta}
e^{-i{k_2}\eta}e^{-i{k_3}\eta}
 \right]+{\rm (perms.)}
 \approx \frac{\sqrt{\epsilon}H^4}{k_3^4k^3}\,,
\label{eq:41}
\end{align}
where the integral is estimated as $O(k_T^{-1})$.
We thus find
\begin{align}
B^{(3)}(k_1,k_2,k_3)\approx \frac{\sqrt{\epsilon}H^4}{k_3^4k^3}\,,
\end{align}
up to a factor of ${\cal O}(1)$. This term is negligible compared
to the leading term given in eq.~\eqref{eq:40} in the sub-curvature
approximation, $k_3\gg 1$.

Finally, we calculate the contribution from
$\mathcal{L}_{int}^{(4)}$. The bispectrum is
\begin{align}
&B^{(4)}(k_1,k_2,k_3)\nnmb
&=
  2\re\left[iu_{k_1}(0)u_{k_2}(0)u_{k_3}(0)
 \left(\int_{-\infty}^0d\eta
 \frac{a^2\dot{\phi}(-k_1^2+k_2^2+k_3^2+1)}{8H(\eta)(k_1^2+4)}
u_{k_1}(\eta)u_{k_2}(\eta)u_{k_3}(\eta)\right)\right.\nnmb
&\qquad\quad+{\rm (perms.)} \Big]\,,
\end{align}
where we have used the property of open harmonics
\begin{align}
 \int d^3{\bf x}\sqrt{\gamma}~Y_{p_1l_1m_1}
\partial_iY_{p_2l_2m_2}\partial^iY_{p_3l_3m_3}
&=\frac{-p_1^2+p_2^2+p_3^2+1}{2} \int d^3{\bf x}\sqrt{\gamma}~Y_{p_1l_1m_1}
Y_{p_2l_2m_2}Y_{p_3l_3m_3}\,.
\end{align}
Using eqs.~\eqref{eq:18}, \eqref{eq:36}, and \eqref{eq:31}, we estimate
the contribution from  $\eta\in(-1,0)$ as
\begin{align}
&\frac{\sqrt{\epsilon}H^4(-k_1^2+k_2^2+k_3^2)}{k_1^5k_2^3k_3^3}
 \re\left[-i\int_{-1}^{0}\frac{d\eta }{\eta^2}
(1+i{k_1}\eta)
(1+i{k_2}\eta)(1+i{k_3}\eta)e^{-i(k_1+k_2+k_3)\eta}
\right]\nnmb
& +{\rm (perms.)} \approx
\frac{\sqrt{\epsilon}H^4}{k_3^4k^3}\,,
\label{eq:27}
\end{align}
where the integral is estimated as $O(k_1k_2k_3/k_T)$.
Using eqs.~\eqref{eq:18}, \eqref{eq:36}, and \eqref{eq:31}, we estimate
the contribution from $\eta\in(-\infty,-1)$ as
\begin{align}
\frac{\sqrt{\epsilon}H^4(-k_1^2+k_2^2+k_3^2)}{k_1^4k_2^2k_3^2}
 \re\left[
\int_{-\infty}^{-1}d\eta e^{\eta}
e^{-i(k_1+k_2+k_3))\eta}
 \right]+{\rm (perms.)}
\approx\frac{\sqrt{\epsilon}H^4}{k_3^4k^3}\,,
\label{eq:23}
\end{align}
where the integral is estimated as $O(k_T^{-1})$. We thus find
\begin{align}
B^{(4)}(k_1,k_2,k_3)\approx \frac{\sqrt{\epsilon}H^4}{k_3^4k^3}\,,
\end{align}
up to a factor of ${\cal O}(1)$. This term is negligible compared
to the leading term given in eq.~\eqref{eq:40} in the sub-curvature
approximation, $k_3\gg 1$.

%%%%%%%%%%%%%%%%%%%%%%%%%%%%%%%%%%%
%%%%%%%%%%%%%%%%%%%%%%%%%%%%%%%%%%%
\section{Open harmonics in the sub-curvature approximation}
\label{sec:sque-limit-mode}
%%%%%%%%%%%%%%%%%%%%%%%%%%%%%%%%%%%
%%%%%%%%%%%%%%%%%%%%%%%%%%%%%%%%%%%

%%%%%%%%%%%%%%%%%%%%%%%%%%%%%%%%%%%
%%%%%%%%%%%%%%%%%%%%%%%%%%%%%%%%%%%
\subsection{Correspondence with flat  harmonics}
\label{sec:openflatharmonics}
%%%%%%%%%%%%%%%%%%%%%%%%%%%%%%%%%%%
%%%%%%%%%%%%%%%%%%%%%%%%%%%%%%%%%%%
In this appendix, we show the correspondence 
between the open and flat harmonics
in the sub-curvature approximation,
where the wavelength of the modes and the separation between points
are assumed to be smaller than the curvature radius, i.e., $1\ll
p$ and $r\ll 1$, respectively. 

On a 3-hyperboloid, whose spatial metric $\gamma_{ij}$ is given by
eq.~\eqref{eq:7}, the
open harmonics $Y_{plm}({\bf x})$ are defined by eq.~\eqref{eq:30}.
On a 3-dimensional flat space, whose metric is given by
\begin{align}
\tilde{\gamma}_{ij}dx^idx^j=dr^2+r^2d\Omega^2_2\,,
\end{align}
the flat spherical harmonics $\tilde{Y}_{klm}({\bf x})$ are defined by \cite{Tanaka:1996cz}
\begin{align}
 \tilde{Y}_{klm}({\bf x})&=\Psi_{kl}(r)Y_{lm}(\Omega),\qquad
 \Psi_{kl}(r) =\sqrt{\frac{2}{\pi}} kj_l(kr)\,,
\label{eq:12}
 \end{align}
where $j_l(x)$ is the spherical Bessel function and
$Y_{lm}(\Omega)$ is the usual spherical harmonics on a 2-sphere.
While $Y_{plm}({\bf x})$ satisfies the relations given by eq.~\eqref{eq:8},
$\tilde{Y}_{klm}({\bf x})$ satisfies the following relations
\begin{align}
\partial^2\tilde{Y}_{klm}({\bf x})&=-k^2\tilde{Y}_{klm}({\bf x})\,,\nnmb
 \tilde{Y}_{klm}^*({\bf x})&=(-1)^m\tilde{Y}_{kl-m}({\bf x})\,,\nnmb
\int d^3{\bf x}\sqrt{\tilde{\gamma}}\, Y_{k_1l_1m_1}^*({\bf x})\tilde{Y}_{k_2l_2m_2}({\bf x})
&=\delta(k_1-k_2)\delta_{l_1l_2}\delta_{m_1m_2}\,,\nnmb
\int_0^\infty dk\sum_{lm}\tilde{Y}_{klm}({\bf x})\tilde{Y}_{klm}({\bf
 x'})&=\delta^{(3)}({\bf x}- {\bf x}')\,. 
\label{eq:73}
\end{align}

\begin{figure}
\begin{center}
 \includegraphics[width=7cm]{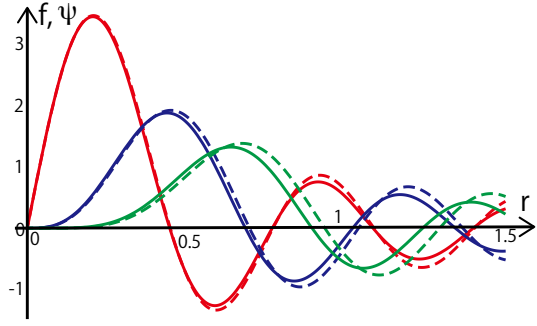}  
\end{center}
\caption{Comparison between $f_{kl}(r)$ (solid lines) and $\Psi_{kl}(r)$
 (dashed lines) for $l=10$ as a function of $r$. The red, blue, and
 green lines show $l=1$, $3$, and $5$, respectively. The two functions
 agree well in $r\ll 1$. 
}
\label{fig:harmonics}
\end{figure}

To begin with, let us find the correspondence between $Y_{plm}({\bf x})$ and
$\tilde{Y}_{klm}({\bf x})$.
As the angular parts $Y_{lm}(\Omega)$ are the same for the open and flat
harmonics, we focus on the correspondence between the
radial parts, i.e., $f_{pl}(r)$ in eq.~\eqref{eq:30} and $\Psi_{kl}(r)$
in eq.~\eqref{eq:12}. 
The correspondence between $f_{pl}(r)$ and $\Psi_{kl}(r)$ with
identification $p\to k$ in the sub-curvature approximation, $1\ll p$ and
$r\ll 1$ is manifest in the differential equations that define the
spherical Bessel functions and the associated Legendre functions
in the sub-curvature approximation.
In figure~\ref{fig:harmonics}, we compare $f_{pl}(r)$ (solid lines) and
$\Psi_{kl}(r)$ (dashed lines) as a function of $r$ for $p=k=10$ and
$l=1$, 3, and 5. 

The asymptotic behaviours of these functions are useful for
understanding their correspondence:
\begin{align}
 \Psi_{kl}(r)\to
\begin{cases}
\ds \sqrt{\frac{2}{\pi}}
\frac{\sin\left(kr-\frac{\pi}{2}l\right)}{r}&(l\ll kr)\\[5pt]
\ds \frac{k^{l+1}r^{l}}{2^{l+1/2}\Gamma(l+3/2)}&(kr\ll l)
\end{cases}\,,
\label{eq:20}
\end{align}
\begin{align}
f_{pl}(r)\to
\begin{cases}
\ds \sqrt{\frac{2}{\pi}}
\frac{\sin\left(pr-\frac{\pi}{2}l\right)}{\sinh r} &(l\ll pr\,,\ 1\ll p)\\[5pt]
\ds \frac{p^{l+1}r^{l}}{2^{l+1/2}\Gamma(l+3/2)}
&(pr\ll l\,,\ 1\ll p)
\end{cases}\,.
\label{eq:32}
\end{align}
In the sub-curvature approximation, where $\sinh r\approx r$, 
$f_{kl}(r)$ and $\Psi_{kl}(r)$ agree with each other
in the two opposite asymptotic regions, $0<r\ll l/k$ and $l/k \ll r \ll 1$.

Next, let us find the correspondence in the square and the cubic integrals of
$Y_{plm}({\bf x})$ and $\tilde{Y}_{klm}({\bf x})$.
The correspondence in the square integrals is exact, as both of them
form an orthonormal set. We write the cubic integrals for $Y_{plm}({\bf
x})$ as 
\begin{equation}
\int d^3{\bf x}\sqrt{\gamma}\,Y_{p_1l_1m_1}({\bf x})Y_{p_2l_2m_2}({\bf x})
Y_{p_3l_3m_3}({\bf x})
=\mathcal{F}_{p_1p_2p_3}^{l_1l_2l_3}\mathcal{G}^{m_1m_2m_3}_{l_1l_2l_3}\,,
\end{equation}
where $\mathcal{F}_{k_1k_2k_3}^{l_1l_2l_3}$ is given in eq.~\eqref{eq:51}
and $\mathcal{G}^{m_1m_2m_3}_{l_1l_2l_3}$ is the Gaunt integral given in
eq.~\eqref{eq:51a}. 
We write the cubic integrals for $\tilde{Y}_{klm}({\bf x})$ as
\begin{equation}
\int d^3{\bf x}\sqrt{\tilde{\gamma}}\,\tilde{Y}_{k_1l_1m_1}({\bf x})\tilde{Y}_{k_2l_2m_2}({\bf x})
\tilde{Y}_{k_3l_3m_3}({\bf x})
=\varPsi_{k_1k_2k_3}^{l_1l_2l_3}\mathcal{G}^{m_1m_2m_3}_{l_1l_2l_3}\,,
\end{equation}
where $\varPsi_{k_1k_2k_3}^{l_1l_2l_3}$ is defined by
\begin{align}
 \varPsi_{k_1k_2k_3}^{l_1l_2l_3}&\equiv
 \int_0^\infty dr r^2 \Psi_{k_1l_1}(r)\Psi_{k_2l_2}(r)\Psi_{k_3l_3}(r)\,.
\label{eq:50}
\end{align}
As the angular parts are given by the Gaunt integral,
$\mathcal{G}_{m_1m_2m_3}^{l_1l_2l_3}$,  we shall focus on the
correspondence between 
$\mathcal{F}_{p_1p_2p_3}^{l_1l_2l_3}$ and $\varPsi_{k_1k_2k_3}^{l_1l_2l_3}$
with the identification of $p_i\to k_i$ for all $i=1$, 2, and 3 in 
the sub-curvature approximation, $1\ll p_i$.

Strictly speaking, however, $1\ll p_i$  is not enough
to show the correspondence in the cubic integral of harmonics.
We thus additionally impose $l_i\ll p_i$ for all $i$
and $1\ll p_{\rm min}$ 
where $p_{\rm min}\equiv\min(|\pm p_1\pm p_2\pm p_3|)$.
In the following, we shall use $k$ not only for the indices of the flat
harmonics, but also for those of the open harmonics. 

Now, there exists an $r^*(\ll 1)$ satisfying 
both $1/k_{\rm min}\ll r^*$ where
$k_{\rm min}\equiv\min(|\pm k_1\pm k_2\pm k_3|)$
and $l_i/k_i \ll r^*$ for all $i$.
Using $r^*$ we can divide the integral in eq.~\eqref{eq:50} into
two parts, corresponding to  $0<r<r^*$ and $r^*<r<\infty$.
In the second part,
by using the $l\ll kr$ case of eq.~\eqref{eq:20}
and the addition theorem for the $\sin$ functions,
we can rewrite the integrand in the form of $\sin[kr+{\rm (phase)}]/r$,
where $k$ is one of $\pm k_1\pm k_2\pm k_3$ for all combinations of $\pm$.
Using the inequality
\begin{align}
 \left|\int_{r^*}^\infty \frac{\sin[kr+{\rm (phase)}]}{r}dr\right|&
=\left|\left[-\frac{\cos [kr+{\rm (phase)}]}{kr}\right]_{r^*}^\infty
+\int_{r^*}^\infty \frac{\cos[kr+{\rm (phase)}]}{kr^2}dr\right|\nnmb
&<\left|\frac{2}{kr^*}\right|\,,
\end{align}
and $1\ll k_{\rm min}r^*$,
we find that the integral over $r^*<r<\infty$ is small
and $\varPsi_{k_1k_2k_3}^{l_1l_2l_3}$ mainly comes from the integral
over $0<r<r^*$. 
Similarly, we find that $\mathcal{F}_{k_1k_2k_3}^{l_1l_2l_3}$ 
mainly comes from the integral over $0<r<r^*$. 
The integrals over $0<r<r^*$ for $\mathcal{F}_{k_1k_2k_3}^{l_1l_2l_3}$ and $\varPsi_{k_1k_2k_3}^{l_1l_2l_3}$
are in fact the same, as the integrands coincide in this region; hence
the correspondence between $\mathcal{F}_{k_1k_2k_3}^{l_1l_2l_3}$ and
$\varPsi_{k_1k_2k_3}^{l_1l_2l_3}$. 
(To see this,  use eq.~\eqref{eq:32} for $pr\ll l$ and
eq.~\eqref{eq:20} for $kr\ll l$, and recall $\sinh r\approx r$ for $r\ll
1$.) In table~\ref{tab:3harmonics}, we show some examples of the
correspondence between $\mathcal{F}_{k_1k_2k_3}^{l_1l_2l_3}$ and
$\varPsi_{k_1k_2k_3}^{l_1l_2l_3}$.

\begin{table}
  \begin{tabular}{c|c|c|c|c|c}
$\begin{pmatrix}l_1&l_2&l_3\\ k_1&k_2&k_3\end{pmatrix}$
&$\begin{pmatrix}1&1&1\\ 3&3&3\end{pmatrix}$ &$\begin{pmatrix}1&1&1\\ 5&8&10\end{pmatrix}$&
$\begin{pmatrix}1&1&1\\ 10&10&10\end{pmatrix}$&$\begin{pmatrix}1&1&3\\ 10&10&10\end{pmatrix}$&
$\begin{pmatrix}1&1&10\\ 10&10&10\end{pmatrix}$\\[10pt] \hline
$\varPsi_{k_1k_2k_3}^{l_1l_2l_3}$ & 0.309  &0.252 & 0.309 &0.190& 0.099\\[5pt] \hline
$\mathcal{F}_{k_1k_2k_3}^{l_1l_2l_3}$& 0.279  &0.246 & 0.306 &0.185& 0.082\\
  \end{tabular}
\caption{Some numerical values of $\varPsi_{k_1k_2k_3}^{l_1l_2l_3}$ and
$\mathcal{F}_{k_1k_2k_3}^{l_1l_2l_3}$. They agree well when $1/k_{\rm
 min}\ll r^*$ (where $k_{\rm min}\equiv\min(|\pm k_1\pm 
 k_2\pm k_3|)$) and $l_i/k_i \ll r^*$ for all $i$ are satisfied.
 }
\label{tab:3harmonics}
\end{table}

%%%%%%%%%%%%%%%%%%%%%%%%%%%%%%%%%%%%%%%%%%%%%%%%%%%%%%%%%%%%%%%
%%%%%%%%%%%%%%%%%%%%%%%%%%%%%%%%%%%%%%%%%%%%%%%%%%%%%%%%%%%%%%%
\subsection{Correspondence with Fourier modes in flat space}
\label{sec:bes_frr}
%%%%%%%%%%%%%%%%%%%%%%%%%%%%%%%%%%%%%%%%%%%%%%%%%%%%%%%%%%%%%%%
%%%%%%%%%%%%%%%%%%%%%%%%%%%%%%%%%%%%%%%%%%%%%%%%%%%%%%%%%%%%%%%
Having established the correspondence between the open and flat
harmonics, let us now establish the correspondence between coefficients
of the open harmonics and those of Fourier expansion in flat space.

Let $\phi({\bf x})$ be a real function in a 3-dimensional flat space.
We shall determine the relation between the multipole expansion
coefficients of the flat harmonics, $\phi_{klm}$, 
and the Fourier coefficients, $\phi\left(\mathbf{k}\right)$.
We expand $\phi({\bf x})$ in two way as
\begin{align}
 \phi\left({\bf x}\right)
&=\int_0^\infty dk \sum_{lm}\Psi_{kl}(r)Y_{lm}(\Omega) \phi_{klm}\,,\\
&=\int \frac{d^3\mathbf{k}}{(2\pi)^3} e^{i\mathbf{k}\mathbf{x}}\phi\left(\mathbf{k}\right)\,,
\end{align}
where  $r=|{\bf x}|$ and $\Omega$ is the direction of ${\bf x}$.
Using eq.~\eqref{eq:12}, reality of $\phi({\bf x})$, and the following identity
\begin{align}
e^{i\mathbf{k}\mathbf{x}}&=\sum_{lm}i^{l} j_l(kr)Y_{lm}^*(\Omega)Y_{lm}(\Omega_{k})\,,
\label{eq:62}
\end{align}
where $k=|{\bf k}|$ and $\Omega_k$ is the direction of ${\bf k}$, we find
\begin{align}
\phi^*_{klm}=
  \frac{i^l}{(2\pi)^{3/2}k}
\int d\Omega_kY_{lm}(\Omega_k)\phi\left(\mathbf{k}\right)\,.
\label{eq:13}
\end{align}
By using eq.~\eqref{eq:12} and eq.~\eqref{eq:62}, we can derive the
useful relation 
\begin{align}
 i^l\frac{k}{(2\pi)^{3/2}}\int d\Omega_{{k}} Y_{lm}(\Omega_{k}) e^{-i\mathbf{k}\mathbf{x}}
% =&i^{l_1}\int d\Omega_{{k}_1} Y_{l_1m_1}(\Omega_{k_1}) 
%\sum_{l'm'}(-i)^{l'} \Psi_{k_1l'}(r)Y_{l'm'}(\Omega)Y_{l'm'}^*(\Omega_{k_1})
=\tilde{Y}_{klm}({\bf x})\,,
\label{eq:14}
\end{align} 
which will be used below.

\subsection{Power spectrum and bispectrum}

We derive the correspondence between the power spectra and
bispectra computed with the Fourier coefficients and the coefficients of
flat harmonics. 
The power spectrum and bispectrum for the Fourier modes are given,
respectively, by
\begin{align}
\big<\phi({{\bf k}_1})\phi({{\bf k}^*_2})\big>= (2\pi)^3\delta({\bf k}_1-{\bf k}_2)P(k_1)\,,
\label{eq:72}
\end{align}
and  
\begin{align}
 \big<\phi({{\bf k}_1})\phi({{\bf k}_2})\phi({{\bf k}_3})\big>=
 (2\pi)^3\delta({\bf k}_1+{\bf k}_2+{\bf k}_3)B(k_1,k_2,k_3)\,.
\label{eq:74}
\end{align}
Using eqs.~\eqref{eq:73}, \eqref{eq:13}, \eqref{eq:14}, \eqref{eq:72}, and
$(2\pi)^3\delta({\bf k}_1-{\bf k}_2)=\int d^3{\bf x}e^{-i({\bf k}_1-{\bf
k}_2){\bf x}}$, we can re-write the power spectrum for the coefficients
of flat harmonics as 
\begin{align}
& \Big<\phi_{k_1l_1m_1}\phi^*_{k_2l_2m_2}\Big> \nnmb
&=  \left(\frac{i^{l_1}k_1}{(2\pi)^{3/2}}
\int d\Omega_kY_{l_1m_1}(\Omega_{k_1})\right)
\left(  \frac{(-i)^{l_2}k_2}{(2\pi)^{3/2}}
\int d\Omega_{k_2}Y_{l_2m_2}^*(\Omega_{k_2})\right)
(2\pi)^3\delta({\bf k}_1-{\bf k}_2)P(k_1)\nnmb
 &=\delta(k_1-k_2)\delta_{l_1l_2}\delta_{m_1m_2}P(k_1)\,.
\label{eq:17}
\end{align}
Similarly, using eqs.~\eqref{eq:73}, \eqref{eq:13}, \eqref{eq:14},
\eqref{eq:74}, and $(2\pi)^3\delta({\bf k}_1+{\bf k}_2+{\bf k}_3)=\int
d^3{\bf x}e^{-i({\bf k}_1+{\bf k}_2+{\bf k}_3){\bf x}}$, 
we can re-write the bispectrum for the coefficients of flat harmonics as
\begin{align}
 \Big<\phi_{k_1l_1m_1}\phi_{k_2l_2m_2}\phi_{k_3l_3m_3}\Big>
%=&i^{l_1+l_2+l_3}\frac{k_1k_2k_3}{(2\pi)^{9/2}}
%\int d\Omega_{k_1} Y_{l_1m_1}^*(\Omega_{k_1}) 
%\int d\Omega_{k_2} Y_{l_2m_2}^*(\Omega_{k_2})
%\int d\Omega_{k_3} Y_{l_3m_3}^*(\Omega_{k_3})  
%(2\pi)^3\delta({\bf k}_1+{\bf k}_2+{\bf k}_3)B(k_1,k_2,k_3)\nnmb
%%=&\int r^2drd\Omega\Psi_{k_1l_1}(r)Y_{l_1m_1}^*(\Omega)
%\Psi_{k_2l_2}(r)Y_{l_2m_2}^*(\Omega)\Psi_{k_3l_3}(r)Y_{l_3m_3}^*(\Omega)
%B(k_1,k_2,k_3)\nnmb
=&\varPsi_{k_1k_2k_3}^{l_1l_2l_3}\mathcal{G}^{m_1m_2m_3}_{l_1l_2l_3}
 B(k_1,k_2,k_3)\,,
\label{eq:131}
\end{align}
where $\varPsi_{k_1k_2k_3}^{l_1l_2l_3}\mathcal{G}^{m_1m_2m_3}_{l_1l_2l_3}$
are defined in eq.~\eqref{eq:50}.

Finally, let us find the correspondence between the power spectra and
bispectra in open and flat universes. 
In an open universe, $P(p_1)$ is defined by
$\big<\phi^*_{p_1l_1m_1}\phi_{p_2l_2m_2}\big>= \delta(p_1-p_2)\delta_{l_1l_2}\delta_{m_1m_2}P(p_1)$,
as in section~\ref{sec:mode_open},
where $\phi_{p_il_im_i}$ are the multipole expansion coefficients with respect to the open harmonics
$Y_{p_il_im_i}$.
Comparing this equation  with eq.~\eqref{eq:17},
we find the correspondence between $P(p_1)$ and $P(k_1)$ with
identification of $p_1\to k_1$. 
Similarly, $B(p_1,p_2,p_3)$ is defined by
$\big<\phi_{p_1l_1m_1}\phi_{p_2l_2m_2}\phi_{p_3l_3m_3}\big>
=B(p_1,p_2,p_3)\mathcal{F}_{p_1p_2p_3}^{l_1l_2l_3}\mathcal{G}^{m_1m_2m_3}_{l_1l_2l_3}$,
as in section~\ref{sec:sque-limit-mode}.
Comparing this equation with eq.~\eqref{eq:131},
and using the correspondence between 
$\mathcal{F}_{p_1p_2p_3}^{l_1l_2l_3}$ and $\varPsi_{k_1k_2k_3}^{l_1l_2l_3}$
in the sub-curvature approximation, we find the correspondence between
$B(p_1,p_2,p_3)$  and $B(k_1,k_2,k_3)$ with the identification of
$p_i\to k_i$.
%%%%%%%%%%%%%%%%%%%%%%%%%
\bibliography{mybib}
\end{document}